\documentclass[aps,prd,onecolumn,floatfix,nofootinbib,showpacs]{revtex4-1}

\pdfoutput=1

\usepackage{graphicx,color,float}
\usepackage{hyperref}
\usepackage{tabularx}
\usepackage{multirow}
\usepackage{verbatim}
\usepackage{longtable}
 \usepackage{amsmath}
\usepackage{amsfonts}
\usepackage{amssymb}
\usepackage{rotating}
\usepackage[dvipsnames]{xcolor}

\usepackage{setspace}

\usepackage{footmisc}

\usepackage{adjustbox}

\usepackage{enumitem}

\newcommand{\imag}{\Im {\rm m}}

\newcommand{\lsim}{\raisebox{-0.13cm}{~\shortstack{$<$ \\[-0.07cm] $\sim$}}~}

\def\slash#1{#1\!\!\!/}

\begin{document}

{\small
\begin{flushright}
IUEP-HEP-25-01
\end{flushright} }

\title{
Higgs boson precision analysis of two-Higgs-doublet models:\\
Full LHC Run 1 and Run 2 data
}

\def\slash#1{#1\!\!/}

\renewcommand{\thefootnote}{\fnsymbol{footnote}}

\author{
\renewcommand{\thefootnote}{\fnsymbol{footnote}}
Yongtae Heo,\footnote{Contact author: yongtae1heo@gmail.com}~
\textsuperscript{\!\!\!\!\!\!,}\footref{corr:author}~
Jae Sik Lee,\footnote{Contact author: jslee@jnu.ac.kr}
and
Chan Beom Park\,\footnote{Contact author: cbpark@jnu.ac.kr}
\textsuperscript{\!\!,}\footnote{\label{corr:author}These authors contributed equally to
this work.}
}

\affiliation{
Department of Physics \& IUEP, Chonnam National University,
Gwangju 61186, Korea
}
\date{April 23, 2025}

\begin{abstract}
\begin{spacing}{1.30}
We present the results obtained by performing global fits
of two-Higgs-doublet models (2HDMs) using
the full Run 1 and Run 2 Higgs datasets collected at the LHC.
Avoiding unwanted
tree-level flavor-changing neutral currents and
including the wrong-sign cases,
we consider 12 scenarios across six types of 2HDMs:
Inert, type I, type II, type III, type IV, and Aligned 2HDMs.
Our main results are presented in
Table~\ref{tab:results} and Fig.~\ref{fig:gof}.
We find that the type-I 2HDM provides the best fit, while
the wrong-sign scenarios of the type-II and type-IV 2HDMs,
where the normalized Yukawa coupling to
down-type quarks is opposite in sign to the Standard Model (SM),
are disfavored.
We also observe that the Aligned 2HDM gives the second-best fit
when the Yukawa couplings to down-type quarks take
the same sign as in the SM,
regardless of the sign of the Yukawa couplings to the charged leptons.
\end{spacing}
\end{abstract}

\maketitle

\renewcommand{\thefootnote}{\arabic{footnote}}

\eject

\section{Introduction}
Since the ATLAS and CMS collaborations independently reported
the observation of a new scalar particle in
the search for the Standard Model (SM) Higgs boson
in 2012~\cite{ATLAS:2012yve,CMS:2012qbp}, its properties
and couplings to SM particles have been extensively
investigated.
At the LHC, the 125 GeV Higgs boson has been observed
through six production processes and via seven decay
modes~\cite{ATLAS:2022vkf,CMS:2022dwd}.
For the Higgs production processes, investigated are
the two main production processes of
gluon-gluon fusion (ggF) and vector-boson fusion (VBF),
along with four subleading ones in which the produced Higgs boson is
associated with a $V = W/Z$ boson (WH/ZH),
a top-quark pair (ttH), or a single top quark (tH).
Through its seven decay modes into $b\bar b$,
$WW^*$, $\tau^+\tau^-$, $ZZ^*$, $\gamma\gamma$, $Z\gamma$,
and $\mu^+\mu^-$,
the couplings to the two massive vector bosons
are well measured with a few percent accuracy.
Third-generation Yukawa couplings are firmly established,
and the decays into
a pair of muons and $Z\gamma$ are now emerging.

\medskip

Recently, we showed that 
the full LHC Higgs precision data are no longer
best described by the SM Higgs boson~\cite{Heo:2024cif}.
The best-fitted values of
the normalized Yukawa couplings are about $2\sigma$ below
their corresponding SM values,
with $1\sigma$ errors of 3\%--5\%
in a model-independent way.
While the deviation of the normalized Yukawa couplings
has been noticed previously in
Refs.~\cite{ATLAS:2022vkf,CMS:2022dwd,Biekotter:2022ckj}
especially when they are universal,
our comprehensive analysis strengthens and confirms this observation
by combining the ATLAS and CMS
Run 2 datasets~\cite{ATLAS:2022vkf,CMS:2022dwd}.\footnote{
The Run 1 dataset~\cite{ATLAS:2016neq} is also included.}

\medskip

In this work, we perform global fits of two-Higgs-doublet models (2HDMs)
without tree-level Higgs-mediated flavor-changing neutral current (FCNC).
The 2HDM models, in which the SM is extended by
adding one more Higgs doublet,
provide one of the most favored and
well-motivated frameworks beyond the SM~\cite{Branco:2011iw}.
Noting that the Aligned 2HDM~\cite{Pich:2009sp} could accommodate
frequently referred 2HDMs without FCNC on the market
by appropriately choosing the model parameters,
we consider six types of CP-conserving 2HDMs in this work,
as summarized in Table~\ref{tab:2hdmtype}.

\medskip

This paper is organized as follows.
Section~\ref{sec:Framework} reviews the Higgs potential of
2HDMs in the Higgs basis and the 125 GeV Higgs couplings
to SM particles under the assumption that the lighter CP-even
neutral state plays the role of the SM Higgs boson.
In Sec.~\ref{sec:AnalysisandResults}, we present the main results
of our analysis, and a detailed discussion of some representative 2HDM
scenarios is given in Sec.~\ref{sec:Discussion}.
Conclusions are drawn in Sec.~\ref{sec:Conclusions}.
In Appendix~\ref{app:A}, we deliver additional details of
the 2HDM scenarios not covered in Sec.~\ref{sec:Discussion}.
%
Appendix~\ref{app:B} shows the correlations among the 125~GeV Higgs boson
couplings to SM particles.
Finally, Appendix~\ref{app:C}
includes
the fitting results obtained by imposing constraints from
the perturbativity of top-quark Yukawa coupling,
the radiative $b\to s\gamma$ decay,
the primary theoretical conditions on the Higgs potential,
and the electroweak precision observables.
%

\section{Framework}
\label{sec:Framework}
In the so-called Higgs basis~\cite{Donoghue:1978cj,Georgi:1978ri}
where only one doublet contains a nonvanishing vacuum expectation value $v$,
the general 2HDM scalar potential can be expressed as~\cite{Lee:2021oaj}
\begin{eqnarray}
\label{eq:VHiggs}
V_{\cal H} &=&
Y_1 ({\cal H}_1^{\dagger} {\cal H}_1)
+Y_2 ({\cal H}_2^{\dagger} {\cal H}_2)
+Y_3 ({\cal H}_1^{\dagger} {\cal H}_2)
+Y_3^{*}({\cal H}_2^{\dagger} {\cal H}_1) \nonumber \\
&&+ Z_1 ({\cal H}_1^{\dagger} {\cal H}_1)^2 + Z_2
({\cal H}_2^{\dagger} {\cal H}_2)^2 + Z_3 ({\cal H}_1^{\dagger}
{\cal H}_1)({\cal H}_2^{\dagger} {\cal H}_2) + Z_4 ({\cal H}_1^{\dagger}
{\cal H}_2)({\cal H}_2^{\dagger} {\cal H}_1) \nonumber \\
&&+ Z_5 ({\cal H}_1^{\dagger} {\cal H}_2)^2 +
Z_5^{*} ({\cal H}_2^{\dagger} {\cal H}_1)^2 + Z_6
({\cal H}_1^{\dagger} {\cal H}_1) ({\cal H}_1^{\dagger} {\cal H}_2) + Z_6^{*}
({\cal H}_1^{\dagger} {\cal H}_1)({\cal H}_2^{\dagger} {\cal H}_1) \nonumber \\
&& + Z_7 ({\cal H}_2^{\dagger} {\cal H}_2) ({\cal H}_1^{\dagger} {\cal H}_2) +
Z_7^{*} ({\cal H}_2^{\dagger} {\cal H}_2) ({\cal H}_2^{\dagger} {\cal H}_1)\;,
\end{eqnarray}
which contains three dimensionful quadratic and seven dimensionless quartic
parameters, four of which are complex.
In this work, we consider the CP-conserving case, assuming that
$\imag(Y_3)=\imag(Z_{5,6,7})=0$.
The complex SU(2)$_L$ doublets of ${\cal H}_1$ and ${\cal H}_2$
can be parametrized as
\begin{eqnarray}
\label{eq:H12inHiggBasis}
{\cal H}_1=
\left(\begin{array}{c}
G^+ \\ \frac{1}{\sqrt{2}}\,(v+\varphi_1+iG^0)
\end{array}\right)\,; \ \ \
{\cal H}_2=
\left(\begin{array}{c}
H^+ \\ \frac{1}{\sqrt{2}}\,(\varphi_2+iA)
\end{array}\right)\,,
\end{eqnarray}
where $v = \left(\sqrt{2}G_F\right)^{-1/2} \simeq 246.22$ GeV, and
$G^{\pm,0}$ and $H^\pm$ stand for the
Goldstone and charged Higgs bosons, respectively.
For the neutral Higgs bosons,
$A$ denotes a CP-odd mass eigenstate and the two states
$\varphi_1$ and $\varphi_2$ mix to form
two CP-even mass eigenstates, one of which should correspond
to the SM Higgs boson.
%
To describe the mixing between the two CP-even states $\varphi_1$ and $\varphi_2$,
we introduce the mixing angle $\gamma$ as follows:\footnote{In the SM limit,
  where $s_\gamma\to 0$, $h=\varphi_1$
plays the role of the SM Higgs boson.}
\begin{eqnarray}
\left(\begin{array}{c}\varphi_1\\ \varphi_2\end{array}\right)=
\left(\begin{array}{cc}
c_\gamma & s_\gamma  \\
-s_\gamma & c_\gamma \end{array}\right)
\left(\begin{array}{c}h\\ H\end{array}\right)\,,
\end{eqnarray}
assuming that $h$ is the lightest neutral Higgs boson with
$M_h\simeq 125$ GeV.
Then, in terms of the four masses $M_{h,H,A,H^\pm}$ and the mixing angle $\gamma$,
the quartic couplings $\{Z_1,Z_4,Z_5,Z_6\}$ are given by
\begin{eqnarray}
\label{eq:Z1456_cpc}
Z_1 &=& \frac{1}{2v^2}\left(c_\gamma^2 M_h^2 + s_\gamma^2 M_H^2 \right)\,, \ \ \
\hspace{1.0cm}
Z_4  =  \frac{1}{v^2}\left( s_\gamma^2 M_h^2 + c_\gamma^2 M_H^2
+M_A^2 -2M_{H^\pm}^2 \right)\,,\nonumber \\[2mm]
Z_5 &=& \frac{1}{2v^2}\left(s_\gamma^2 M_h^2
+c_\gamma^2 M_H^2 -M_A^2 \right)\,, \ \ \
Z_6  =  \frac{1}{v^2}\left(-M_h^2 + M_H^2 \right)c_\gamma s_\gamma\,.
\end{eqnarray}
To summarize, the CP-conserving 2HDM scalar potential is fully specified
by fixing the 9 elements of the following input-parameter set
\begin{equation}
{\cal I}^{V_{\cal H}}=\left\{v\,,M_h\,;
\gamma\,,M_H\,,M_A\,,M_{H^\pm}\,;Z_2\,,Z_3\,,Z_7\right\}\,,
\end{equation}
taking account of the tadpole conditions of
$Y_1  +  Z_1 v^2 = 0$ and $Y_3  + Z_6 v^2/2 = 0$, and
using the relation $Y_2=M_{H^\pm}^2-Z_3 v^2/2$.

\medskip

\begin{table}[!t]
\caption{\label{tab:2hdmtype}
The alignment parameters of the 2HDMs considered in this work.
For the four (I, II, III, IV) types of 2HDM
based on the Glashow-Weinberg condition~\cite{Glashow:1976nt},
we follow the conventions found in,
for example, Ref.~\cite{Lee:2021oaj}.
}
\setlength{\tabcolsep}{0.5ex}
\renewcommand{\arraystretch}{1.3}
\begin{center}
\begin{tabular}{|l|c|c|c|c|c|c|}
\hline
 & ~~~~~Inert~~~~~ & I   & II
& III & IV & Aligned \\
\hline
$\zeta_u$ & $0$ & $1/t_\beta$ & $1/t_\beta$ & $1/t_\beta$ & $1/t_\beta$ & $\zeta_u$
\\
$\zeta_d$ & $0$ & $1/t_\beta$ & $-t_\beta$ & $1/t_\beta$ & $-t_\beta$  & $\zeta_d$ \\
$\zeta_\ell$ & $0$ & $1/t_\beta$ & $-t_\beta$ & $-t_\beta$ & $1/t_\beta$  &
$\zeta_\ell$ \\
\hline
& &
$\zeta_d=\zeta_\ell=\zeta_u$ &
$\zeta_d=\zeta_\ell=-1/\zeta_u$ &
$\zeta_d=-1/\zeta_\ell=\zeta_u$ &
$\zeta_d=-1/\zeta_\ell=-1/\zeta_u$ & all independent \\
\hline
\end{tabular}
\end{center}
\end{table}
To perform global fits to the full Higgs datasets collected
at the LHC and Tevatron, we adopt the conventions and notations
of the $h$ couplings to SM particles following
Ref.~\cite{Choi:2021nql}.
The Yukawa couplings of $h$ are given by:
\begin{equation}
\label{eq:hff}
{\cal L}_{h\bar{f}f}\ =\ - \sum_{f=\{u,d,\ell\}}\,g^S_{h\bar{f}f}\,\frac{m_f}{v}
\left( \bar{f}\,f\right)\, h\,,
\end{equation}
where $\{u,d,\ell\}$ collectively represents the up-type quarks, down-type quarks,
and charged leptons. 
The normalized coupling to the scalar fermion bilinear $\bar{f}f$
is given by:
\begin{equation}
g^S_{h\bar{f}f} \ = \ c_\gamma-\zeta_f s_\gamma\,.
\end{equation}
Note that we have introduced three alignment parameters
$\zeta_{u,d,\ell}$ as in the Aligned 2HDM~\cite{Pich:2009sp},
recognizing that all the 2HDMs
considered in this work can be accommodated by appropriately choosing the three
align parameters, as detailed in Table~\ref{tab:2hdmtype}.
%
Additionally, we assume $\zeta_u > 0$ without loss of generality, exploiting the
rephasing invariance of the Higgs potential and the Yukawa
interactions~\cite{Lee:2021oaj}.
The couplings of $h$ to the massive vector bosons are described by:
\begin{equation}
{\cal L}_{hVV}  =  g\,M_W \, \left( g_{_{hWW}}\,W^+_\mu W^{- \mu}\ + \
g_{_{hZZ}}\,\frac{1}{2c_W^2}\,Z_\mu Z^\mu\right) \, h\,,
\end{equation}
with
\begin{equation}
g_{_{hWW}}=g_{_{hZZ}}\equiv g_{_{hVV}}=c_\gamma\,.
\end{equation}
For the loop-induced $h$ couplings to $gg$, $\gamma\gamma$, and $Z\gamma$,
taking $M_h=125$ GeV, we have~\cite{Heo:2024cif}\footnote{For the normalizations of the form factors
$S^{g\,,\gamma\,,Z\gamma}$ and their general expressions,
and also for the loop functions $F_0$ and $I_1$ below,
we refer to Ref.~\cite{Choi:2021nql}.
The SM values for these form factors are:
$S^g_{\rm SM}=0.636 + 0.071\,i$,
$S^\gamma_{\rm SM}=-6.542 + 0.046\,i$, and
$S^{Z\gamma}_{\rm SM} =-11.6701 + 0.0114 \, i$.
}
\begin{eqnarray}
\label{eq:loop_induced}
S^g&=&  0.688\,g_{h\bar{t}t}^S + (-0.043+0.063\,i)\,g_{h\bar{b}b}^S
+  (-0.009+0.008\,i)\,g_{h\bar{c}c}^S\,,\nonumber \\[2mm]
S^\gamma&=& -8.324\,g_{_{hWW}} + 1.826\,g_{h\bar{t}t}^S
+(-0.020 + 0.025\,i)\,g_{h\bar{b}b}^S
   +(-0.024 + 0.022\,i)\,g_{h\tau\tau}^S + \Delta S^{\gamma}\,, \nonumber\\[2mm]
S^{Z\gamma} &=&
        -12.3401  \, g_{_{hWW}}
        +  0.6891  \, g_{h\bar{t}t}^S
        + (-0.0186 + 0.0111 \, i ) \, g_{h\bar{b}b}^S
        + (-0.0005 + 0.0002 \, i) \,g_{h\tau\tau}^S + \Delta S^{Z\gamma} \,,
\end{eqnarray}
with
\begin{equation}
\Delta S^\gamma=+ g_{_{hH^+H^-}}\frac{v^2}{2 M_{H^\pm}^2}
F_0(\tau_{hH^\pm})\,, \ \
\Delta S^{Z\gamma}= +2\, \frac{g_{_{hH^+H^-}}}{c_Ws_W}\,
\frac{v^2}{M_{H^\pm}^2}\,
I_1\left(\tau_{hH^\pm},\lambda_{H^\pm}\right)\,,
\end{equation}
where $\tau_{hH^\pm}=M_h^2/4M_{H^\pm}^2$,
$\lambda_{H^\pm}=M_{Z}^2/4M_{H^\pm}^2$, and
\begin{equation}
g_{_{h H^+ H^-}} =
c_\gamma\,  Z_3-s_\gamma\,  Z_7 \,.
\end{equation}
When $M_{H^\pm}\gg M_{h,Z}$, we find
\begin{equation}
\label{eq:za_a}
\frac{\Delta S^{Z\gamma}\,(H^\pm)}{\Delta S^\gamma\,(H^\pm)} \simeq
\frac{4}{c_Ws_W}\,\frac{I_1(0,0)}{F_0(0)} =
\frac{4}{c_Ws_W}\,\frac{1/6}{1/3} \simeq 4.8\,,
\end{equation}
and
\begin{equation}
\Delta S^\gamma\,(H^\pm) \simeq \frac{1}{6}\,\left[g_{_{hH^+H^-}}\,
\left(\frac{v}{M_{H^\pm}}\right)^2\right]\,.
\end{equation}
Note that the form factors $\Delta S^\gamma$ and 
$\Delta S^{Z\gamma}$ are strongly correlated in 2HDMs.
This is contrasted to the model-independent approach 
taken in Ref.~\cite{Heo:2024cif},
where they are completely independent.
To summarize, the $h$ couplings to SM particles
are fully specified
by fixing the 5 elements of the following input-parameter set:
\begin{equation}
{\cal I}^{\rm Couplings}
=\left\{\zeta_u\,,\zeta_d\,,\zeta_\ell\,;
g_{_{hH^+H^-}}\,,M_{H^\pm}\right\}\,,
\end{equation}
in addition to the mixing angle $\gamma$ appearing in the Higgs potential parameter
set ${\cal I}^{V_{\cal H}}$.

\medskip

For our numerical analysis,
we adopt the basis in which $\zeta_u \geq 0$ and organize the input parameters
for the Higgs potential and the $h$ couplings as follows:
\begin{eqnarray}
{\cal I}^{V_{\cal H}} \oplus
{\cal I}^{\rm Couplings} &=&
\bigg\{v,M_h\bigg\}_2 \oplus
\bigg\{\gamma,M_{H^\pm},g_{_{hH^+H^-}};\zeta_u,\zeta_d,\zeta_\ell\bigg\}_{3,4,6}
\oplus \bigg\{M_H,M_A,Z_2,Z_7\bigg\}_4\,,
\end{eqnarray}
with $Z_3 =(g_{_{h H^+ H^-}} +s_\gamma\,  Z_7)/c_\gamma$.
The middle set,
which contains 3 (Inert), 4 (I, II, III, IV), or 6 (Aligned) free parameters,
is directly relevant for our Higgs boson precision analysis of 2HDMs.
As in Ref.~\cite{Heo:2024cif},
depending on the 2HDM type,
we use the following short notations for the $h$ couplings
in our global fits:
\begin{eqnarray}
&&
C_V=g_{_{hVV}}\,;
C_u^S=g^S_{h u\bar{u}}\,,
C_d^S=g^S_{h d\bar{d}}\,,
C_\ell^S=g^S_{h \ell^-\ell^+}\,({\rm Aligned})\,;
C_{Vf}=g_{_{hVV}}=g^S_{h u\bar{u}}=g^S_{h d\bar{d}}=g^S_{h \ell^-\ell^+}\,({\rm Inert})\,;
\nonumber\\[2mm]
&&
C_f^S=g^S_{h u\bar{u}}=g^S_{h d\bar{d}}=g^S_{h \ell^-\ell^+}\,({\rm I})\,,
C_{d\ell}^S=g^S_{h d\bar{d}}=g^S_{h \ell^-\ell^+}\,({\rm II})\,,
C_{ud}^S=g^S_{h u\bar{u}}=g^S_{h d\bar{d}}\,({\rm III})\,,
C_{u\ell}^S=g^S_{h u\bar{u}}=g^S_{h \ell^-\ell^+}\,({\rm IV})\,.
\end{eqnarray}

\section{Analysis and Results}
\label{sec:AnalysisandResults}
To perform global fits of the $h$ couplings to 
the Tevatron and full LHC Run 1 and Run 2 Higgs
datasets, we
use a total of 77 experimental signal strengths: 76 of them are taken from
Refs.~\cite{tevatron_aa_ww,Herner:2016woc,ATLAS:2016neq,ATLAS:2022vkf,CMS:2022dwd}
and presented in Tables I, II, III, and IV of Ref.~\cite{Heo:2024cif}
and the 77th one
comes from the $h\to Z\gamma$ data,
obtained from the combined ATLAS and CMS
analysis~\cite{ATLAS:2023yqk}.\footnote{Specifically, we use $\mu^{\rm
    EXP}(pp\to h\to Z\gamma)=\widehat\mu^{\rm EXP}(Z\gamma)=2.2\pm 0.7$, see Appendix E of Ref.~\cite{Heo:2024cif}.}
For the LHC Run 1~\cite{ATLAS:2016neq}
and Run 2~\cite{ATLAS:2022vkf,CMS:2022dwd} datasets,
we take account of correlation among the experimental signal strengths
within each data set.
On the other hand,
we refer to Sec.~III.B of Ref.~\cite{Heo:2024cif}
for the details of the theoretical signal strength calculations.
These calculations assume that each theoretical signal strength is given by the
product of the production and decay signal strengths:
$\mu({\cal P},{\cal D}) \simeq\widehat\mu({\cal P})\ \widehat\mu({\cal D})$.
Once all the theoretical signal strengths $\mu_i$'s
associated with specific production processes and
decay modes are obtained, we use the $\chi^2$ statistic for $n$ correlated observables:
\begin{equation}
\chi^2_n = \sum^n_{i,j=1}
\frac{(\mu_i-\mu^{\rm EXP}_i)}{\sigma^{\rm EXP}_i}\,
\left(\rho^{-1}\right)_{ij}\,
\frac{(\mu_j-\mu^{\rm EXP}_j)}{\sigma^{\rm EXP}_j}\,,
\end{equation}
where $i$, $j$ index the $n$ correlated production-times-decay modes and $\rho$ is
the $n \times n$ correlation matrix.
Specifically, for the LHC Run 1, we use the correlation matrix 
given in Fig. 27 of Ref.~\cite{ATLAS:2016neq}.
For those of the LHC Run 2, 
see Auxiliary Fig.~14 presented in 
{\bf https://doi.org/10.17182/hepdata.130266} (ATLAS) and
the figure entitled
{\it ``Production times decay signal strength modifiers correlations"}
provided in the website
{\bf https://dx.doi.org/10.17182/hepdata.127765} (CMS).
%
For our chi-square analysis, we also consider the goodness of fit (gof), which
quantifies the agreement with the experimentally measured signal strengths in a
given fit.
Note that the gof approaches 1 as the value of $\chi^2$ per degree
of freedom (dof) becomes smaller.

\begin{table}[!hbt]
\centering
\caption{\label{tab:scenarios}
The 12 scenarios considered in our global fit to the six types of 2HDMs and
the varying parameters in each scenario. Also shown are
the novel combinations which are well constrained by the Higgs
precision data.
}  \vspace{1mm}
\renewcommand{\arraystretch}{1.3}
\begin{tabular}{c||c||c||c|c}
\hline
Types & Scenarios
& Varying Parameters & \multicolumn{2}{c}{Novel Combinations Constrained}  \\ \hline
Inert & {\bf Inrt}  & $s_\gamma\,,g_{_{hH^+H^-}}\,,M_{H^\pm}$ &
$s_\gamma$ &
\multirow{12}{*}{$g_{_{hH^+H^-}}\left(\frac{v}{M_{H^\pm}}\right)^2$}
\\ \cline{1-4}
Type I & {\bf I}  &
\multirow{7}{*}{$s_\gamma\,,t_\beta=\frac{1}{\zeta_u}\,,g_{_{hH^+H^-}}\,,M_{H^\pm}$} &
$-s_\gamma/t_\beta$   &
\\ \cline{1-2} \cline{4-4}
\multirow{2}{*}{Type II} &
{\bf II$^+$} $(C_{d\ell}^S>0)$ &  &
$s_\gamma(t_\beta-1/t_\beta)$   &
\\ \cline{2-2} \cline{4-4}
&
{\bf II$^-$} $(C_{d\ell}^S<0)$ &  & $s_\gamma t_\beta$ & \\ \cline{1-2} \cline{4-4}
\multirow{2}{*}{Type III} & {\bf III$^+$} $(C_{\ell}^S>0)$ &   & $s_\gamma(t_\beta-1/t_\beta)$ &
\\ \cline{2-2} \cline{4-4}
&
{\bf III$^-$} $(C_{\ell}^S<0)$ &  & $s_\gamma t_\beta$ &  \\ \cline{1-2} \cline{4-4}
\multirow{2}{*}{Type IV} &
{\bf IV$^+$} $(C_{d}^S>0)$ &   & $-s_\gamma/t_\beta$ & \\ \cline{2-2} \cline{4-4}
&
{\bf IV$^-$} $(C_{d}^S<0)$ &  & $s_\gamma t_\beta$ &  \\ \cline{1-4} \cline{4-4}
\multirow{4}{*}{Aligned} &
{\bf A$^{++}$} $(C_d^S>0\,,C_{\ell}^S>0)$ &
\multirow{4}{*}{$s_\gamma\,,\zeta_u\,,\zeta_d\,,\zeta_\ell\,,
g_{_{hH^+H^-}}\,,M_{H^\pm}$} &
\multirow{4}{*}{$-s_\gamma\zeta_u\,,-s_\gamma\zeta_d\,,-s_\gamma\zeta_\ell$} &
\\ \cline{2-2}
&
{\bf A$^{+-}$} $(C_d^S>0\,,C_{\ell}^S<0)$ & &  \\ \cline{2-2}
&
{\bf A$^{-+}$} $(C_d^S<0\,,C_{\ell}^S>0)$ & &  \\ \cline{2-2}
&
{\bf A$^{--}$} $(C_d^S<0\,,C_{\ell}^S<0)$ & &  \\ \hline
%
\end{tabular}
\end{table}

\medskip

The so-called wrong-sign alignment can occur when $s_\gamma \zeta_f = 2$ with
$c_\gamma\to 1$,
resulting in normalized Yukawa couplings that are equal in strength but opposite
in sign to the SM couplings. Since $c_\gamma = g_{_{hVV}}$ is constrained to be
(very) close to 1 by precision Higgs data, the wrong-sign alignment can occur
only when $|\zeta_f|\gg 1$.
In our global fits of the six types of 2HDMs, we treat the wrong-sign case as an
independent scenario, as it occupies a completely different region of parameter
space in the $(s_\gamma, \zeta_f)$ plane compared to the same-sign case, where
$|s_\gamma\zeta_f|$ is small.
We observe that $C_{d\ell}^S$, $C_\ell^S$, and $C_d^S$
can take the wrong signs in type II, III, and IV 2HDMs, respectively.
Meanwhile, in the Aligned 2HDM,
$C_d^S$ and $C_\ell^S$ are independent of each other and from $C_u^S$.
As a result, we have identified 12 distinct scenarios, as shown in
Table~\ref{tab:scenarios}. In the same table,
we also list the parameters to be varied in each scenario and
the novel combinations
which turn out to be well constrained by the Higgs precision data.

%
\begin{table}[!htb]
\caption{\label{tab:results}
The minimal chi-square per degree of freedom ($\chi^2_{\rm min}$/dof),
goodness of fit (gof),
$1\sigma$ confidence interval of $s_\gamma$, and the best-fitted values
of the  125 GeV Higgs couplings to SM particles
in the 12 scenarios considered in our global fit to the six types of 2HDMs.
For the SM, we obtain
$\chi^2_{\rm min}/{\rm dof}=85.29/77$ and ${\rm gof}=0.2424$.
}  \vspace{1mm}
\renewcommand{\arraystretch}{1.3}
\begin{adjustbox}{width=\textwidth}
$\begin{array}{|c|c||c|c||c||c|cccccc|c|c|}
\hline
\multicolumn{2}{|c||}{}& \chi^2_\textrm{min}/\textrm{dof} & \textrm{gof} & s_\gamma & C_{V} & \multicolumn{6}{c|}{C_{f}^{S} } & \Delta S^{\gamma} & \Delta S^{Z\gamma}  \\ \hline
%
\multirow{2}{*}{\textbf{Inrt}} & s_\gamma>0 &
\multirow{2}{*}{80.28/74} & \multirow{2}{*}{0.2889} & [0, 0.25] &
\multicolumn{7}{c|}{\multirow{2}{*}{ $C_{Vf}= 0.9872^{+0.0128}_{-0.0198}$ }} &
\multirow{2}{*}{$-0.421^{+0.180}_{-0.200}$} & \multirow{2}{*}{$-2.100^{+0.938}_{-0.910}$} \\ \cline{2-2} \cline{5-5}
 & s_\gamma<0  & & & [-0.25, 0] & \multicolumn{7}{c|}{} & &
\\ \hline
\multicolumn{2}{|c||}{\textbf{I} } & 75.94/73 & 0.3839 & [0, 0.18] & 0.9999^{+0.0001}_{-0.0166} & \multicolumn{6}{c|}{ 0.929^{+0.033}_{-0.029} } & -0.216^{+0.179}_{-0.200} & -1.036^{+0.859}_{-1.020}  \\ \hline
\multirow{2}{*}{$\textbf{II}^+$} & s_\gamma>0 & 79.82/73 & 0.2734 & [0, 0.02] & 1.0_{-0.0001} & \multicolumn{3}{c|}{ C_{u}^{S} = 0.971^{+0.028}_{-0.025} } & \multicolumn{3}{c|}{ C_{d\ell}^S = 1.0^{+0.010}_{-0.000} } & -0.413^{+0.176}_{-0.182} & -1.993^{+0.853}_{-0.871}  \\ \cline{2-14}
& s_\gamma<0 & 79.85/73 & 0.2727 & [-0.02, 0] & 1.0_{-0.0001} & \multicolumn{3}{c|}{ C_{u}^{S} = 1.0^{+0.008}_{-0.000} } & \multicolumn{3}{c|}{ C_{d\ell}^S = 0.969^{+0.031}_{-0.029} } & -0.261^{+0.206}_{-0.193} & -1.254^{+0.988}_{-0.940}  \\ \hline
\multicolumn{2}{|c||}{\textbf{II}^- } & 86.62/73 & 0.1317 & [-0.25, -0.02] & 0.9968^{+0.0030}_{-0.0281} & \multicolumn{3}{c|}{ C_{u}^{S} = 1.0^{+0.001}_{-0.000} } & \multicolumn{3}{c|}{ C_{d\ell}^S = -0.993^{+0.043}_{-0.034} } & -0.350^{+0.212}_{-0.209} & -1.682^{+1.018}_{-1.007}  \\ \hline
\multirow{2}{*}{$ \textbf{III}^+ $} & s_\gamma>0 & 79.11/73 & 0.2923 & [0, 0.02] &  1.0_{-0.0003} & \multicolumn{3}{c|}{ C_{ud}^S = 0.951^{+0.037}_{-0.036} } & \multicolumn{3}{c|}{ C_{\ell}^{S} = 1.0^{+0.010}_{-0.000} } & -0.300^{+0.174}_{-0.180} & -1.443^{+0.839}_{-0.866}  \\ \cline{2-14}
& s_\gamma<0 & 79.10/73 & 0.2923 & [-0.02, 0] &  1.0_{-0.0003} & \multicolumn{3}{c|}{ C_{ud}^S = 1.0^{+0.011}_{-0.000} } & \multicolumn{3}{c|}{ C_{\ell}^{S} = 0.950^{+0.037}_{-0.039} } & -0.331^{+0.175}_{-0.172} & -1.593^{+0.843}_{-0.822}  \\ \hline
\multicolumn{2}{|c||}{\textbf{III}^- } & 78.96/73 & 0.2962 & [-0.15, -0.02] &  0.9998^{+0.00002}_{-0.0110} & \multicolumn{3}{c|}{ C_{ud}^S = 1.0^{+0.0003}_{-0.0000} } & \multicolumn{3}{c|}{ C_{\ell}^{S} = -0.951^{+0.040}_{-0.037} } & -0.374^{+0.172}_{-0.176} & -1.800^{+0.831}_{-0.855}  \\ \hline
\multicolumn{2}{|c||}{\textbf{IV}^+ } & 78.60/73 & 0.3062 & [0, 0.02] &  1.0_{-0.0002} & \multicolumn{3}{c|}{ C_{u\ell}^S = 0.968^{+0.021}_{-0.020} } & \multicolumn{3}{c|}{ C_{d}^{S} = 1.0^{+0.015}_{-0.000} } & -0.397^{+0.174}_{-0.176} & -1.907^{+0.837}_{-0.848}  \\ \hline
\multicolumn{2}{|c||}{\textbf{IV}^- } & 85.55/73 & 0.1495 & [-0.15, -0.02] &  0.9997^{+0.00005}_{-0.0110} & \multicolumn{3}{c|}{ C_{u\ell}^S = 1.0^{+0.0000}_{-0.0002} } & \multicolumn{3}{c|}{ C_{d}^{S} = -1.035^{+0.031}_{-0.040} } & -0.402^{+0.185}_{-0.213} & -1.955^{+0.909}_{-1.004}  \\ \hline
\multicolumn{6}{c|}{} &\multicolumn{2}{c|}{ C_{u}^{S} } &\multicolumn{2}{c|}{ C_{d}^{S} } & \multicolumn{2}{c|}{ C_{\ell}^{S} } & \multicolumn{2}{c}{ }  \\ \hline
\multicolumn{2}{|c||}{\textbf{A}^{++} } & 75.70/71 & 0.3302 & [0, 0.30] &  0.9970^{+0.0030}_{-0.0413} &\multicolumn{2}{c|}{ 0.931^{+0.036}_{-0.048} } &\multicolumn{2}{c|}{ 0.912^{+0.048}_{-0.095} } & \multicolumn{2}{c|}{ 0.923^{+0.039}_{-0.053} } & -0.182^{+0.221}_{-0.213} & -0.875^{+1.064}_{-1.025} \\ \hline
\multicolumn{2}{|c||}{\textbf{A}^{+-} } & 75.52/71 & 0.3347 & [0.02, 0.30] &  0.9918^{+0.0080}_{-0.0366} &\multicolumn{2}{c|}{ 0.928^{+0.038}_{-0.043} } &\multicolumn{2}{c|}{ 0.899^{+0.060}_{-0.084} } & \multicolumn{2}{c|}{ -0.917^{+0.046}_{-0.044} } & -0.219^{+0.212}_{-0.215} & -1.051^{+1.018}_{-1.037} \\ \hline
\multicolumn{2}{|c||}{\textbf{A}^{-+} } & 77.20/71 & 0.2873 & [0.02, 0.28] &  0.9993^{+0.0005}_{-0.0391} &\multicolumn{2}{c|}{ 0.888^{+0.039}_{-0.037} } &\multicolumn{2}{c|}{ -0.910^{+0.086}_{-0.052} } & \multicolumn{2}{c|}{ 0.930^{+0.036}_{-0.054} } & -0.167^{+0.213}_{-0.220} & -0.803^{+1.023}_{-1.079} \\ \hline
\multicolumn{2}{|c||}{\textbf{A}^{--} } & 77.03/71 & 0.2918 & [0.02, 0.28] &  0.9931^{+0.0067}_{-0.0333} &\multicolumn{2}{c|}{ 0.888^{+0.040}_{-0.037} } &\multicolumn{2}{c|}{ -0.894^{+0.072}_{-0.068} } & \multicolumn{2}{c|}{ -0.921^{+0.044}_{-0.047} } & -0.200^{+0.194}_{-0.230} & -0.961^{+0.934}_{-1.103} \\ \hline
\end{array}$
\end{adjustbox}
\end{table}
\begin{figure}[!htb]
\begin{center}
\includegraphics[width=16.5cm]{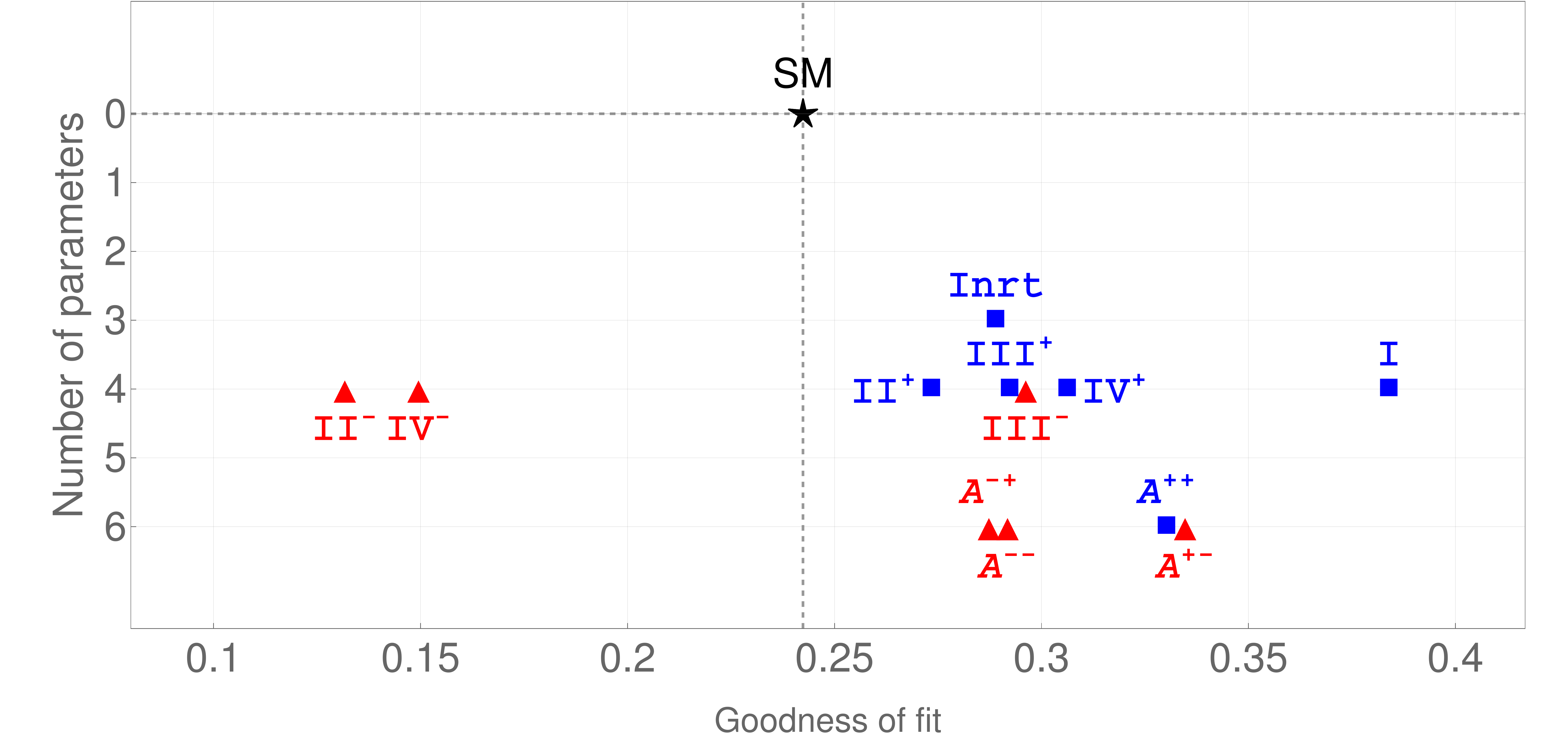}
\end{center}
\vspace{-0.5cm}
\caption{\it
Goodness of fit for the
12 scenarios considered in our global fit of the six types of 2HDMs.
Blue boxes represent the scenarios in which all the Yukawa couplings are positive,
while the wrong-sign scenarios are denoted by red triangles.
The SM point is marked with a star.
}
\label{fig:gof}
\end{figure}

In Table~\ref{tab:results}, we present our main results
obtained by performing global fits of 2HDMs
using the full Higgs datasets collected at the LHC and Tevatron.
In Fig.~\ref{fig:gof}, we compare the gof values for
the 12 scenarios of six types of 2HDMs.
The current Higgs precision data pull the SM gof down to $0.2424$. 
This is basically due to the deviations of the combined signal strengths of 
the Higgs decay modes into $\gamma\gamma$, $bb$, 
and $\tau\tau$ from the SM: see Table VI 
and Sec.~IV.A.2 in Ref.~\cite{Heo:2024cif}.
In {\bf Inrt},
two degenerate minima appear depending on the sign of $s_\gamma$, although the
Higgs couplings cannot distinguish the sign.
We also find
(almost) degenerate minima in the {\bf II}$^+$ and {\bf III}$^+$ scenarios, depending
on the sign of $s_\gamma$.
However, 
we conclude 
that it is not a parametric feature, but rather an
accidental one
since we observe that
this degeneracy is lifted when turning off the correlation in the LHC
datasets.
Scenario {\bf I} provides the best fit among all the 2HDMs
considered in this work.
In fact, all the 2HDM scenarios
except {\bf II}$^-$ and {\bf IV}$^-$
yield the better gof values than the SM one.
We also observe that the wrong-sign $C_d^S$ results in worse fits
in the {\bf A}$^{-\pm}$ scenario compared to {\bf A}$^{+\pm}$.
This is due to the interference between the top- and bottom-quark contributions
to ggF, which leads to larger values of $\chi^2$ when $C_d^S$
takes the wrong sign.
On the other hand, when only $C_\ell^S$ takes the wrong sign as
in {\bf III}$^-$, {\bf A}$^{+-}$, and {\bf A}$^{--}$,
the fits are very slightly better compared to the corresponding scenarios with
positive $C_\ell^S$. 
This improvement is because the flip of sign of $C_\ell^S$ 
induces a shift in $\Delta S^\gamma$,
leading to a larger deviation
of $\Delta S^{Z\gamma}$ from 0 which improves the fit to the $h\to Z\gamma$ data,
see Eqs.~(\ref{eq:loop_induced}) and (\ref{eq:za_a}).
The preference for the wrong-sign $C_\ell^S$ is a new feature driven by the
inclusion of the $h\to Z\gamma$ data, although it is nearly meaningless with
the current statistical precision.

\medskip

For the couplings, our finding are as follows:
\begin{itemize}
\item
In {\bf Inrt}, the gauge-Higgs coupling $C_V$ is equal to
the normalized Yukawa coupling $C_f^S$, and $C_{Vf}$ is consistent with the SM,
with $1\sigma$ errors of 1\%--2\%.
\item
The $1\sigma$ confidence intervals (CIs) of $s_\gamma$ in {\bf II}$^+$,
{\bf III}$^+$, and {\bf IV}$^+$ are at a few percent level.
The deviation of $C_V=c_\gamma$ from 1 is extremely small, with
$1\sigma$ errors of $0.01$\%--$0.03$\%.
In other cases,
the best-fitted values of $C_V$ are very close to 1,
with $1\sigma$ errors of 1\%--4\%.
\item
In {\bf I} for which we have obtained the best value of gof,
the best-fitted value of
the normalized Yukawa coupling is about $2\sigma$ below
the SM value of 1, with $1\sigma$ errors of 3\%.
\item
In the wrong-sign scenarios of {\bf II}$^-$, {\bf III}$^-$, and {\bf IV}$^-$,
the best-fitted values of $C_{d\ell}^S$, $C_\ell^S$ and
$C_d^S$ are around $-1$ with $1\sigma$ errors of 4\%.
The best-fitted values for the other couplings, $C_u^S$ ({\bf II}$^-$),
$C_{ud}^S$ ({\bf III}$^-$), and $C_{u\ell}^S$ ({\bf IV}$^-$), are
extremely close to 1, with $1\sigma$ errors of less than $0.1$\%.
\item
In {\bf II}$^+$ and {\bf III}$^+$, each of which has
two (almost) degenerate minima,
the best-fitted values are:
$(C_u^S\,,C_{d\ell}^S)\simeq(0.97\,,1)$ and
$(C_{ud}^S\,,C_{\ell}^S)\simeq(0.95\,,1)$
when $s_\gamma>0$ and $t_\beta\ll 1$.\footnote{Recall that
$C_{u\,,ud\,,u\ell}^S=1-s_\gamma/t_\beta$ and
$C_{d\ell\,,\ell\,,d}^S=1+s_\gamma\,t_\beta$.}
Conversely, when $s_\gamma<0$ and $t_\beta\gg 1$,
they become $(C_u^S\,,C_{d\ell}^S)\simeq(1\,,0.97)$ and
$(C_{ud}^S\,,C_{\ell}^S)\simeq(1\,,0.95)$.
In {\bf IV}$^+$, the best-fitted values are
$(C_{u\ell}^S\,,C_{d}^S)\simeq(0.97\,,1)$.
The $1\sigma$ errors are about 1\% (2\%--4\%)
when the best-fitted value is equal to (smaller than) 1.
\item
In each of {\bf II}$^\pm$, {\bf III}$^\pm$, and {\bf IV}$^\pm$,
the best-fitted value of one of the Yukawa couplings is close to 1, while
the other is
approximately
$0.97\,(0.95)$ or $-1\,(-0.95)$.
%
\item
In {\bf A}$^{\pm\pm,\pm\mp}$,
the best-fitted values of $C_V$ are very close to 1,
with $1\sigma$ errors of about 4\%.
The best-fitted values of
the normalized Yukawa coupling are about $2\sigma$ below
the SM value of 1,
with upper $1\sigma$ errors of about 3\%--5\%
when they are positive.
When negative, they are around
$-0.9$, with $1\sigma$ errors of 4\%--8\%.
\item
The loop-induced coupling $\Delta S^\gamma$, arising
from triangle loops involving the charged Higgs bosons,
is generally consistent with the SM in
{\bf I}, {\bf II}$^{+}\,(s_\gamma<0)$,  and {\bf A}$^{\pm\pm,\pm\mp}$
with $1\sigma$ errors of about 0.2.
In other cases,
$\Delta S^\gamma$ is about $2\sigma$ below the SM value of 0,
with $1\sigma$ errors of around 0.2.
For $\Delta S^{Z\gamma}$, we note that
the relation $\Delta S^{Z\gamma} \simeq 5\, \Delta S^\gamma$
holds, as seen in Eq.~(\ref{eq:za_a}).
\end{itemize}

\section{Discussion}
\label{sec:Discussion}
In this Section, we provide a detailed description of the results from
our global fits of the 2HDMs in the scenarios
{\bf Inrt}, {\bf I}, {\bf II}$^\pm$, and {\bf A}$^{++}$.
For the other scenarios, we refer to Appendix~\ref{app:A}.
\begin{figure}[!htb]
\begin{center}
\includegraphics[width=16.5cm]{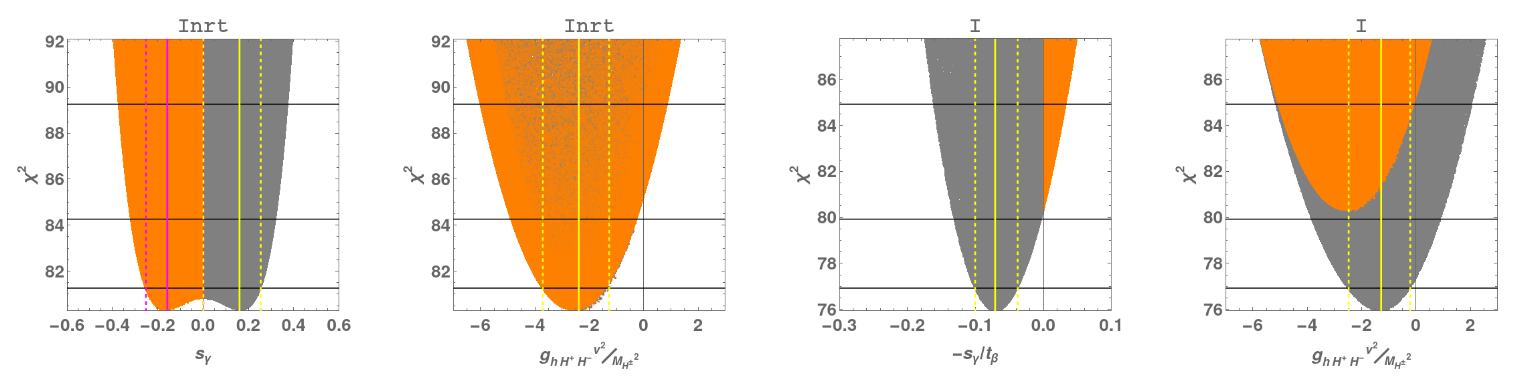}
\end{center}
\vspace{-0.5cm}
\caption{\it {\bf Inrt} [Two Left] and {\bf I} [Two Right]:
$\chi^2$ above the minimum
versus, from left to right,
$-s_\gamma$,
$g_{_{hH^+H^-}}(v^2/M_{H^\pm}^2)$,
$-s_\gamma/t_\beta$, and
$g_{_{hH^+H^-}}(v^2/M_{H^\pm}^2)$
with the orange and gray shaded regions
for $s_\gamma<0$ and $s_\gamma>0$, respectively.
The horizontal lines correspond to $\Delta\chi^2=1$, $4$, $9$, while
the magenta and yellow vertical lines indicate the minima and
the $1\sigma$ CIs for $s_\gamma<0$ and $s_\gamma>0$, respectively.
}
\label{fig:t01}
\end{figure}

\medskip

In the two left frames of Fig.~\ref{fig:t01}, we show $\chi^2$ above the minimum
as a function of
$s_\gamma$ and the combination $g_{_{hH^+H^-}}(v^2/M_{H^\pm}^2)$ in the {\bf Inrt} scenario.
Note that, in the frame for $\chi^2$ versus $g_{_{hH^+H^-}}(v^2/M_{H^\pm}^2)$,
the orange ($s_\gamma<0$) and gray ($s_\gamma>0$) shaded regions completely overlap.
We obtain $\chi^2_{\rm min}/{\rm dof}=80.28/74$ and ${\rm gof}=0.2889$.
Two degenerate minima are found around $s_\gamma = \pm 0.2$, with the $1\sigma$
CIs of $[-0.25\,,0]$ for $s_\gamma<0$ and $[0\,,0.25]$ for
$s_\gamma>0$, which leads to $C_{Vf}=0.9872^{+0.0128}_{-0.0198}$.
For $g_{_{hH^+H^-}}(v^2/M_{H^\pm}^2)$,
the minimum occurs around $-2.5$, with the $1\sigma$ CI of $[-3.7\,,-1.3]$,
independent of the sign of $s_\gamma$. This results in
$\Delta S^\gamma =-0.421^{+0.180}_{-0.200}$ and
$\Delta S^{Z\gamma} =-2.100^{+0.938}_{-0.910}\simeq 5\, \Delta S^\gamma$.
Note that, for {\bf Inrt}, it is not possible to determine the sign of
$s_\gamma$ through Higgs boson precision analysis.

\medskip

In the two right frames of Fig.~\ref{fig:t01}, we show $\chi^2$ above the
minimum as a function of
the two novel combinations, $-s_\gamma/t_\beta$ and
$g_{_{hH^+H^-}}(v^2/M_{H^\pm}^2)$, in the {\bf I} scenario.
We obtain $\chi^2_{\rm min}/{\rm dof}=75.94/73$ and ${\rm gof}=0.3839$.
We have obtained the best gof for the {\bf I} scenario.
The $1\sigma$ CIs for the first and second novel
combinations are $[-0.10\,,-0.04]$ and $[-2.5\,,-0.2]$, respectively, while the
$1\sigma$ CI for $s_\gamma$ is $[0\,,0.18]$.
For the couplings, we find
$C_V=0.9999^{+0.0001}_{-0.0166}$, $C_f^S=0.929^{+0.033}_{-0.029}$,
$\Delta S^\gamma =-0.216^{+0.179}_{-0.200}$ and
$\Delta S^{Z\gamma} =-1.036^{+0.859}_{-1.020}$,
which are consistent with the $1\sigma$ CIs.

\medskip

In the scenario of {\bf II}$^{+}$,
in addition to $g_{_{hH^+H^-}}(v^2/M_{H^\pm}^2)$ as in {\bf Inrt} and {\bf I},
we introduce the following novel combination:
\begin{equation}
s_\gamma\left(t_\beta-\frac{1}{t_\beta}\right)\,,
\end{equation}
which accounts for the deviation of the normalized Yukawa couplings
from the SM value of 1 in the limit $c_\gamma \to 1$ when $t_\beta$
is very large or very small.
In this limit,
either $s_\gamma t_\beta$ or $s_\gamma/t_\beta$ remains finite,
despite $c_\gamma \to 1$.
We use this same combination of
$s_\gamma(t_\beta-1/t_\beta)$ to analyze the
Higgs precision data in {\bf III}$^{+}$, as shown in Table~\ref{tab:scenarios}.
We find that this combination is useful for addressing the accidental degeneracy
in the {\bf II}$^{+}$ and {\bf III}$^{+}$ scenarios.

\begin{figure}[!t]
\begin{center}
\includegraphics[width=14.0cm]{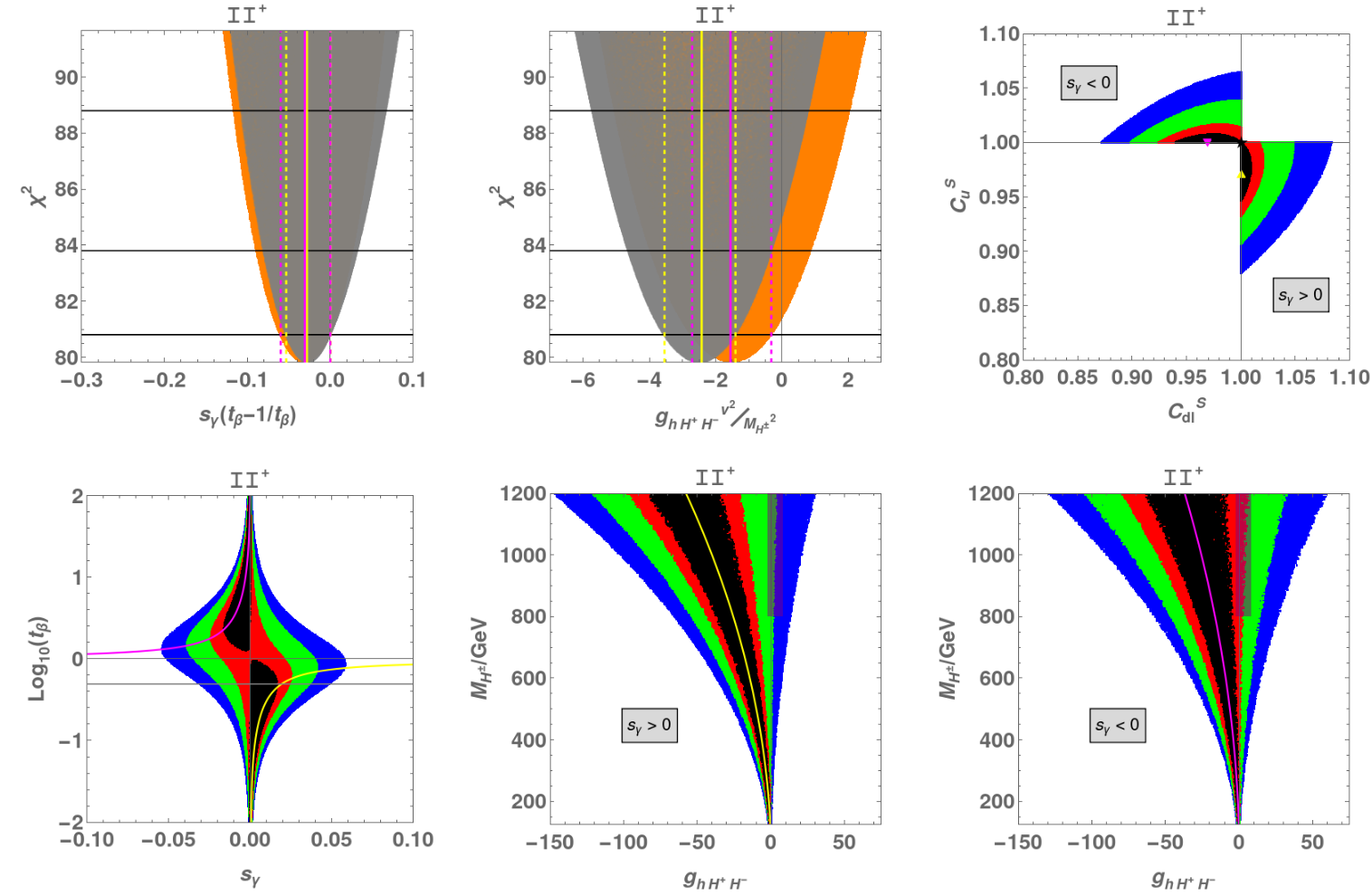}
\end{center}
\vspace{-0.5cm}
\caption{\it {\bf II$^+$}:
[Upper]
$\chi^2$ above the minimum versus
$s_\gamma(t_\beta-1/t_\beta)$ (left) and
$g_{_{hH^+H^-}}(v^2/M_{H^\pm}^2)$ (middle)
and the CL regions in
the $(C_{d\ell}^S\,,C_u^S)$ plane (right).
In the left and middle frames, the orange and gray shaded regions
represent $s_\gamma<0$ and $s_\gamma>0$, respectively.
The horizontal lines correspond to $\Delta\chi^2=1$, $4$, $9$, while
the magenta and yellow vertical lines indicate the minima and
the $1\sigma$ CIs for $s_\gamma<0$ and $s_\gamma>0$, respectively.
[Lower]
The CL regions in the $(s_\gamma\,,\log_{10}t_\beta)$ plane (left)
and in the $(g_{_{hH^+H^-}}\,,M_{H^\pm})$ planes (middle and right).
The magenta ($s_\gamma<0$) and yellow ($s_\gamma>0$)
curves represent the functional relations
between the two parameters at the minima given by
$s_\gamma(t_\beta-1/t_\beta)\simeq -0.03$ (left) and
$g_{_{hH^+H^-}}(v^2/M_{H^\pm}^2)\simeq
-2.5\,({\rm middle}\,;s_\gamma>0)\,,-1.5\,({\rm right}\,;s_\gamma<0)$.
In the left frame, the horizontal line
denotes the lower limit of $t_\beta>1/2$,  and
the lightly shaded narrow regions
in the middle and right frames
correspond to the constraints of
$-2.5 < g_{_{h H^+ H^-}} < 8.1$ and $M_{H^\pm}>800\,{\rm GeV}$.
In the upper-right and lower frames, the contour regions represent
$\Delta\chi^2\leq 1$ (black),
$\Delta\chi^2\leq 2.3$ (red),
$\Delta\chi^2\leq 5.99$ (green),
$\Delta\chi^2\leq 11.83$ (blue)
above the minimum, which correspond to
confidence levels of 39.35\%, 68.27\%, 95\%, and 99.73\%, respectively.
}
\label{fig:t2p}
\end{figure}
\medskip
In the upper-left and upper-middle frames of Fig.~\ref{fig:t2p},
we show $\chi^2$ above the minimum
as a function of the novel combinations for {\bf II$^+$}.
We obtain $\chi^2_{\rm min}/{\rm dof} = 79.82/73$ and $79.85/73$, with ${\rm
  gof} = 0.2734$ and $0.2727$, respectively.
For {\bf II$^+$},
we find two nearly degenerate minima around $s_\gamma(t_\beta - 1/t_\beta) =
-0.03$. For example, when $s_\gamma < 0$, $(s_\gamma\,, t_\beta) \simeq (-0.01\,,
3)$, and when $s_\gamma > 0$, $(s_\gamma\,, t_\beta) \simeq (0.01\,, 1/3)$.
This accidental degeneracy is a
characteristic feature of the Run 2 LHC Higgs datasets.
%
The two nearly degenerate minima for $s_\gamma(t_\beta-1/t_\beta)$
are quite close to each other, with a $1\sigma$ CI of
$[-0.06\,,0]$, without distinguishing them.
On the other hand, they are slightly separated for
$g_{_{hH^+H^-}}(v^2/M_{H^\pm}^2)$, with $1\sigma$ CIs of
$[-2.7\,,-0.3]$ for $s_\gamma < 0$ (orange) and $[-3.6\,,-1.4]$ for $s_\gamma > 0$
(gray).
Incidentally, the $1\sigma$ CIs for $s_\gamma$ are
$[-0.02\,,0]$ for $s_\gamma < 0$ and $[0\,,0.02]$ for $s_\gamma > 0$.
See Table~\ref{tab:results} for the couplings, which are consistent
with the $1\sigma$ CIs of $s_\gamma$, $s_\gamma(t_\beta-1/t_\beta)$, and
$g_{_{hH^+H^-}}(v^2/M_{H^\pm}^2)$.
Note that the two degenerate minima of {\bf II}$^+$ are clearly
separated in the $(C_u^S\,,C_{d\ell}^S)$ plane.
See the upper-right frame of Fig.~\ref{fig:t2p}
for the confidence level (CL) regions in the plane,
where the two minima are denoted by triangles.
Note that both of $C_u^S$ and $C_{d\ell}^S$ cannot be larger or
smaller than 1 simultaneously because
$C_{u}^S=1-s_\gamma/t_\beta$ and
$C_{d\ell}^S=1+s_\gamma\,t_\beta$.
When $s_\gamma>0$,
the minimum occurs for large $t_\beta$, with
$s_\gamma(t_\beta-1/t_\beta)\simeq s_\gamma t_\beta$, and
it is located at $(C_u^S\,,C_{d\ell}^S) \simeq (0.97\,,1)$.
When $s_\gamma<0$, on the other hand,
the minimum occurs for small $t_\beta$, with
$s_\gamma(t_\beta-1/t_\beta)\simeq -s_\gamma/t_\beta$, and
it is located at $(C_u^S\,,C_{d\ell}^S) \simeq (1\,,0.97)$.

\medskip
In the three lower frames of Fig.~\ref{fig:t2p}, we show
the CL regions in the
$(s_\gamma\,,\log_{10}t_\beta)$ and
$(g_{_{hH^+H^-}}\,,M_{H^\pm})$ planes for {\bf II}$^+$
where the magenta ($s_\gamma<0$) and yellow ($s_\gamma>0$)
curves denote the functional relation
between the two parameters at the minima given by
$s_\gamma(t_\beta-1/t_\beta)\simeq -0.03$ (left) and
$g_{_{hH^+H^-}}(v^2/M_{H^\pm}^2)\simeq
-2.5\,({\rm middle})\,,-1.5\,({\rm right})$,
as seen in the upper-left and upper-middle frames.
We observe that the $\chi^2$ behavior of the two
novel combinations describes the CL regions in the factored planes
quite well.

\medskip

The alignment parameter $\zeta_u=1/t_\beta$ cannot be
significantly larger than 1 since a large $\zeta_u$
(or equivalently, a small $t_\beta$) leads to a
nonperturbative top-quark Yukawa coupling and a Landau pole
near the TeV scale.
On the other hand,
the analysis of the radiative $b\to s\gamma$ decay
within the type-II and type-IV 2HDMs yields the 95\% CL constraint of
$M_{H^\pm}>800$ GeV~\cite{Misiak:2020vlo}.
In addition, the absolute values of the quartic couplings $Z_{i=1-7}$
cannot be arbitrarily large
if the  perturbative unitarity (UNIT) conditions and
those for the Higgs potential to be bounded from below (BFB)
are imposed. By combining the UNIT and BFB conditions with
the electroweak (ELW) constraint at 95\% CL, the quartic couplings
$Z_3$ and $Z_7$ are found to be restricted as
$-2.4 \lsim Z_3 \lsim 8.0$ and $-2.7\lsim Z_7\lsim 2.7$~\cite{Lee:2021oaj},
which,
with $Z_3 =(g_{_{h H^+ H^-}} +s_\gamma\,  Z_7)/c_\gamma$, might lead to
$-2.5 \lsim g_{_{h H^+ H^-}} \lsim 8.1$ in {\bf II}$^+$.
In lower frames of Fig.~\ref{fig:t2p},
the horizontal line (left)
indicates the lower limit of $t_\beta>1/2$,
and the lightly shaded narrow regions (middle and right)
represent
the constraints of
$-2.5 < g_{_{h H^+ H^-}} < 8.1$ and $M_{H^\pm}>800\,{\rm GeV}$.
While the constraint of $t_\beta>1/2$ does not significantly affect the fitting results,
we find that when $M_{H^\pm}>800$ GeV, the  
allowed region obtained by collectively
imposing the UNIT, BFB, and ELW (UBE) constraints
lies outside the $68\,(39)$\% CL region for $s_\gamma>0$ ($s_\gamma<0$)
in the $(g_{_{h H^+ H^-}}\,, M_{H^\pm})$ plane.
Note that,
if the UBE constraints are imposed for $M_{H^\pm}>800$ GeV, the degeneracy is lifted,
and the minimum occurs when $s_\gamma<0$.
Taking all the constraints into account,
we
obtain $(\chi^2_{\rm min})_{\rm Constrained}^{{\bf II}^+}/{\rm dof}=81.12/73$ and
$({\rm gof})_{\rm Constrained}^{{\bf II}^+}=0.2410$.
In this work, we present the unconstrained fitting results,
presented in Table~\ref{tab:results}, as our main ones
to avoid theoretical biases, while
keeping in mind the possibility of
the 2HDM framework could be viewed as a low-energy effective field theory
like as the SM itself.
This approach ensures that the regions of parameter space of phenomenological
interest are fully captured without theoretical prejudices or indirect
experimental constraints coined from assuming a specific model.
%
Otherwise, see Table~\ref{tab:results_constrained}
for the fitting results obtained by imposing
$t_\beta>1/2$ ({\bf I}, {\bf II}$^\pm$, {\bf III}$^\pm$, {\bf IV}$^\pm$)
or $\zeta_u<2$ (Aligned), the UBE constraints, and $M_{H^\pm}>800$ GeV.
The last constraint on $M_{H^\pm}$ is applied only
in {\bf II}$^\pm$ and {\bf IV}$^\pm$.
%

\begin{figure}[!t]
\begin{center}
\includegraphics[width=18.5cm]{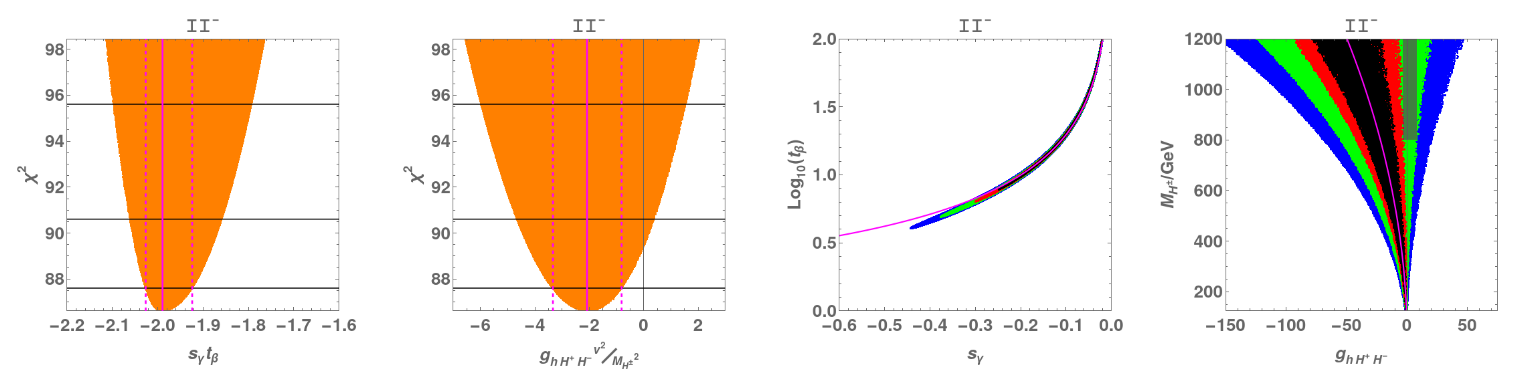}
\end{center}
\vspace{-0.5cm}
\caption{\it {\bf II}$^-$:
$\chi^2$ above the minimum versus
$s_\gamma t_\beta$ and
$g_{_{hH^+H^-}}(v^2/M_{H^\pm}^2)$ (two left)
and the CL regions in the $(s_\gamma\,,\log_{10}t_\beta)$
and $(g_{_{hH^+H^-}}\,,M_{H^\pm})$ planes (two right).
The lines, shades, and colors are the same as in Fig.~\ref{fig:t2p}.
}
\label{fig:t2m}
\end{figure}
\medskip
In the two left frames of Fig.~\ref{fig:t2m}, we show $\chi^2$ above the minimum
versus the novel combinations for {\bf II$^-$}.
We obtain $\chi^2_{\rm min}/{\rm dof}=86.62/73$
and ${\rm gof}=0.1317$, which represents the worst fit among the 12 scenarios.
We note that the minimum occurs
around $s_\gamma t_\beta\simeq -2$
when $s_\gamma<0$ and $t_\beta\gg 1$
with its $1\sigma$ CI of $[-2.03\,,-1.92]$.
The $1\sigma$ CI for $g_{_{hH^+H^-}}(v^2/M_{H^\pm}^2)$ is $[-3.3\,, -0.8]$, and for
$s_\gamma$, it is $[-0.25\,, -0.02]$.
Note that, in this scenario,
while $C_{d\ell}^S\simeq -1$ with a $1\sigma$ error of 4\%,
$C_u^S$ is very close to the SM value of 1, with
the $1\sigma$ errors of less than $0.1$\%.
Compared to {\bf II}$^+$,
the range of $s_\gamma$ is about 10 times broader,
reaching to $\sim -0.4$ at 95\% CL.
This is illustrated in the middle-right frame, where the magenta curve denotes
the functional relation between $s_\gamma$ and $t_\beta$ at the minima,
given by $s_\gamma t_\beta \simeq -2$.
%
In the right frame,
the magenta curve corresponds to
$g_{_{hH^+H^-}}(v^2/M_{H^\pm}^2)\simeq -2$, and
the lightly shaded narrow region represents the constraints
imposed by the UBE constraints and $M_{H^\pm}>800$ GeV.
When $M_{H^\pm}>800$ GeV, the UBE
allowed region $-3.1\lsim g_{_{hH^+H^-}}\lsim 8.4$
lies just outside the $68$\% CL region.
Under these conditions, we obtain
$(\chi^2_{\rm min})_{\rm Constrained}^{{\bf II}^-}/{\rm dof}=89.07/73$ and
$({\rm gof})_{\rm Constrained}^{{\bf II}^-}=0.0972$.
Note that $s_\gamma>0$ is completely ruled out in {\bf II}$^{-}$.

\medskip

\begin{figure}[!htb]
\begin{center}
\includegraphics[width=18.5cm]{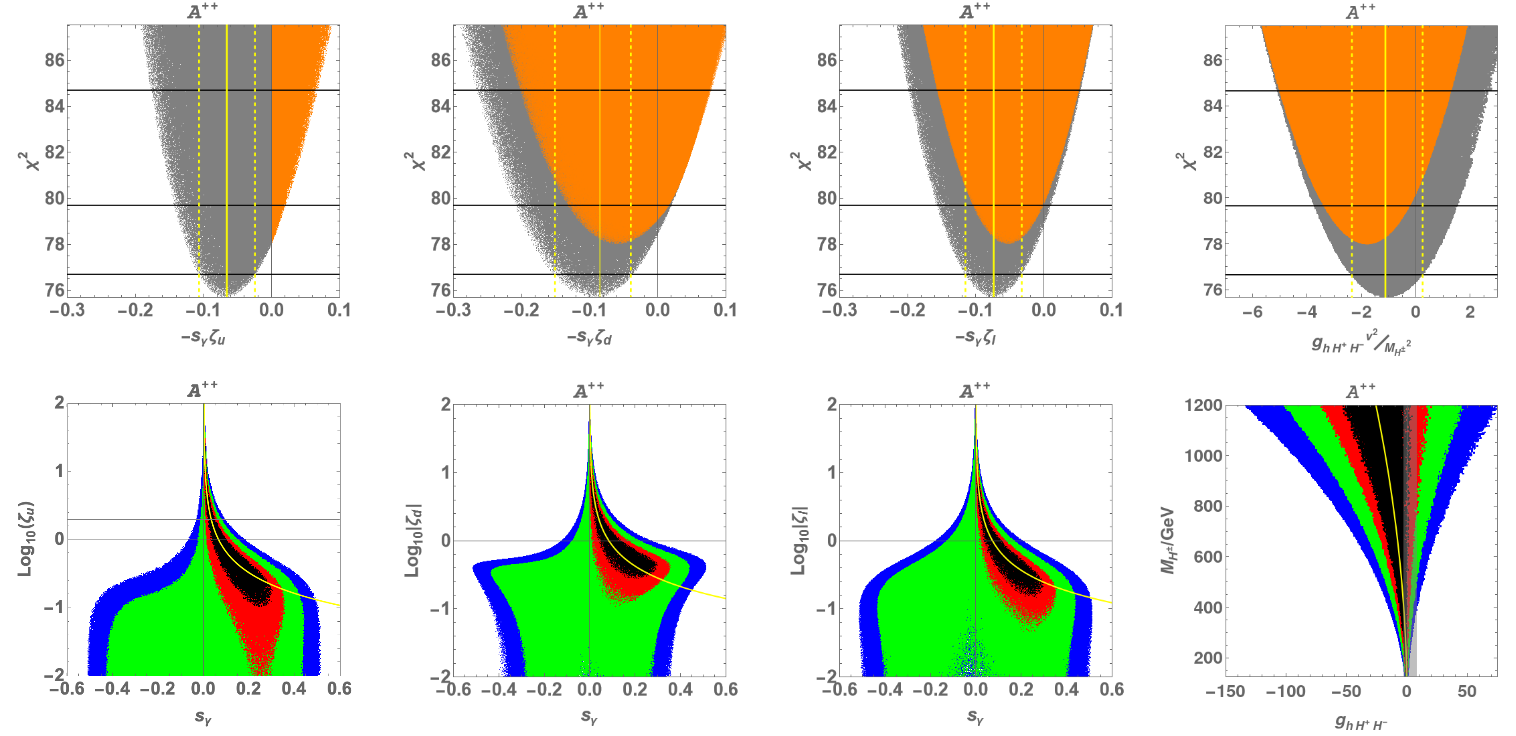}
\end{center}
\vspace{-0.5cm}
\caption{\it {\bf A}$^{++}$:
[Upper] From left to right, $\chi^2$ above the minimum
versus
$-s_\gamma\zeta_u$,
$-s_\gamma\zeta_d$,
$-s_\gamma\zeta_\ell$, and
$g_{_{hH^+H^-}}(v^2/M_{H^\pm}^2)$.
[Lower] From left to right, the CL regions in the
$(s_\gamma\,,\log_{10}\zeta_u)$,
$(s_\gamma\,,\log_{10}|\zeta_d|)$,
$(s_\gamma\,,\log_{10}|\zeta_\ell|)$, and
$(g_{_{hH^+H^-}}\,,M_{H^\pm})$ planes.
The lines, shades, and colors are the same as in Fig.~\ref{fig:t2p},
but in the lower-left frame, the horizontal line
denotes the upper limit of $\zeta_u<2$.
}
\label{fig:app}
\end{figure}

In the upper four frames of Fig.~\ref{fig:app},
we show $\chi^2$ above the minimum
versus the novel combinations of
$-s_\gamma\zeta_u$,
$-s_\gamma\zeta_d$,
$-s_\gamma\zeta_\ell$, and
$g_{_{hH^+H^-}}(v^2/M_{H^\pm}^2)$ for {\bf A}$^{++}$.
We obtain $\chi^2_{\rm min}/{\rm dof}=75.67/71$
and ${\rm gof}=0.3302$ constituting
the second-best fits together with {\bf A}$^{+-}$, see Fig.~\ref{fig:gof}.
For {\bf A}$^{++}$, we find the minima around
$-s_\gamma\zeta_u=-0.06$,
$-s_\gamma\zeta_d=-0.09$,
$-s_\gamma\zeta_\ell=-0.07$, and
$g_{_{hH^+H^-}}(v^2/M_{H^\pm}^2)=-1$, with $1\sigma$ CIs of
$[-0.10,-0.03]$, $[-0.15,-0.04]$, $[-0.12,-0.04]$, and
$[-2.13,+0.21]$, respectively.
We observe that the positions of the minima and the $1\sigma$ CIs
match well with the best-fitted values presented
in Table~\ref{tab:results}.
Notably, the minima occur when $s_\gamma>0$
(gray shaded regions) and, accordingly, $\zeta_{u,d,\ell}>0$.
In the four lower frames of Fig.~\ref{fig:app}, we show
the CL regions in the
$(s_\gamma\,,\log_{10}\zeta_u)$,
$(s_\gamma\,,\log_{10}|\zeta_d|)$,
$(s_\gamma\,,\log_{10}|\zeta_\ell|)$, and
$(g_{_{hH^+H^-}}\,,M_{H^\pm})$ planes,
where the yellow 
curves denote the functional relation
between the two parameters at the minima, given by
$s_\gamma\zeta_u\simeq 0.06$,
$s_\gamma\zeta_d\simeq 0.09$,
$s_\gamma\zeta_\ell\simeq 0.07$, and
$g_{_{hH^+H^-}}(v^2/M_{H^\pm}^2)\simeq -1$.
Once again, we observe that the $\chi^2$ behavior of these four novel
combinations adequately describes the CL regions in the factored planes.
By imposing $t_\beta>1/2$, the UBE conditions, and $M_{H^\pm}>800$ GeV,  we obtain
$(\chi^2_{\rm min})_{\rm Constrained}^{{\bf A}^{++}}/{\rm dof}=75.99/73$ and
$({\rm gof})_{\rm Constrained}^{{\bf A}^{++}}=0.3211$.

\medskip

We refer to Appendix~\ref{app:B}
for a comprehensive view of the CL regions in the two-coupling planes, showing
all possible correlations among the 125 GeV Higgs boson couplings to SM
particles in the 12 scenarios of the six types of 2HDMs analyzed in this work.

\section{Conclusions}
\label{sec:Conclusions}
We perform global fits of 2HDMs to
the full Run 1 and Run 2 Higgs datasets collected at the LHC,
with the integrated luminosities per experiment of approximately
5~fb$^{-1}$ at 7~TeV, 20~fb$^{-1}$ at 8~TeV~\cite{ATLAS:2016neq},
and up to 139~fb$^{-1}$ at 13~TeV~\cite{ATLAS:2022vkf,CMS:2022dwd}.
For the $H\to Z\gamma$ signal strength, we use the most recent result
obtained in the combined ATLAS and CMS analysis~\cite{ATLAS:2023yqk}.
Requiring the absence of tree-level Higgs-mediated FCNCs,
we consider 12 scenarios across six types of 2HDMs:
Inert, type I, type II, type III, type IV, and Aligned 2HDMs.
The wrong-sign cases are treated as independent scenarios.
We assume that the lightest neutral Higgs state plays
the role of the 125 GeV Higgs boson discovered at the LHC
and focus on its couplings to SM particles.
The fitting results are presented in
Table~\ref{tab:results} and Fig.~\ref{fig:gof}.

\medskip

We find that the type-I 2HDM provides the best fit,
with $\chi^2_{\rm min}/{\rm dof}=75.94/73$ and
${\rm gof}=0.3839$, compared to the SM values of
$(\chi^2_{\rm min}/{\rm dof})_{\rm SM}=85.29/77$ and
$({\rm gof})_{\rm SM}=0.2424$.
In this case, all three normalized Yukawa
couplings are identical\footnote{
In the type-II, type-III, and type-IV 2HDMs,
unlike in the type-I 2HDM,
the normalized Yukawa coupling to up-type quarks and
that to down-type fermions cannot simultaneously be larger or
smaller than 1.
This explains why these models result in worse fits
to the current LHC Higgs datasets.},
with the best-fitted value about $2\sigma$ below
the SM value of 1, and $1\sigma$ errors of 3\%.
On the other hand, the coupling to the two massive vector bosons
saturates to the SM value of 1,
with a lower $1\sigma$ error of 2\%.
The loop-induced couplings to $\gamma\gamma$
and $Z\gamma$, when normalized to their corresponding SM values,
deviate from the SM value of 0 by about $1\sigma$,
with $1\sigma$ errors of 3\% and 9\%, respectively.

\medskip

We observe that the Aligned 2HDM provides the second-best fit
when the Yukawa couplings to down-type quarks take
the same sign as in the SM, regardless of
the sign of the Yukawa couplings to the charged leptons.\footnote{In fact,
  we observe slightly better fits
  when only the Yukawa coupling to the charged leptons has the wrong sign.}
On the other hand, only the wrong-sign scenarios
of type-II and type-IV 2HDMs result in
worse fits than the SM among the 12 scenarios considered in this work.
Otherwise,
the Inert, type-II (with $C_{d\ell}^S>0)$,
type-III, type-IV (with $C_{d}^S>0$),  wrong-sign Aligned (with $C_d^S<0$)
2HDMs form the third-best fit group.

\medskip

Last but not least, when the UBE constraints are imposed
along with $M_{H^\pm}>800$ GeV in the type-II and type-IV 2HDMs,
we find worse fits than the SM, even when the signs of the Yukawa
are the same as in the SM, see Fig.~\ref{fig:gof_constrained}.

%
%
\section*{Acknowledgment}
C.B.P. gratefully acknowledges the hospitality at APCTP during the focus program
``Dark Matter as a Portal to New Physics'' (APCTP-2025-F01).
This work was supported by the National Research Foundation (NRF) of Korea
Grant No. NRF-2021R1A2B5B02087078 (Y.H., J.S.L.).
The work of C.B.P. was supported by the
NRF grant funded by
the Korea government (Ministry of Science and ICT) (No. RS-2023-00209974).
The work was also supported in part by
the NRF of Korea Grants No. NRF-2022R1A5A1030700 and No. RS-2024-00442775.


\def\theequation{\Alph{section}.\arabic{equation}}
\begin{appendix}
\setcounter{equation}{0}
\section{Scenarios of {\bf III}$^\pm$, {\bf IV}$^\pm$,
{\bf A}$^{+-}$, {\bf A}$^{-+}$, and {\bf A}$^{--}$}
\label{app:A}
In this Appendix, we present $\chi^2$ above the minimum versus
the novel combinations and the CL regions in
the factored two-dimensional planes for
the scenarios of {\bf III}$^\pm$, {\bf IV}$^\pm$,
{\bf A}$^{+-}$, {\bf A}$^{-+}$, and {\bf A}$^{--}$, which
were not discussed in Sec.~\ref{sec:Discussion}.

%

\medskip

\begin{figure}[!htb]
\begin{center}
\includegraphics[width=14.0cm]{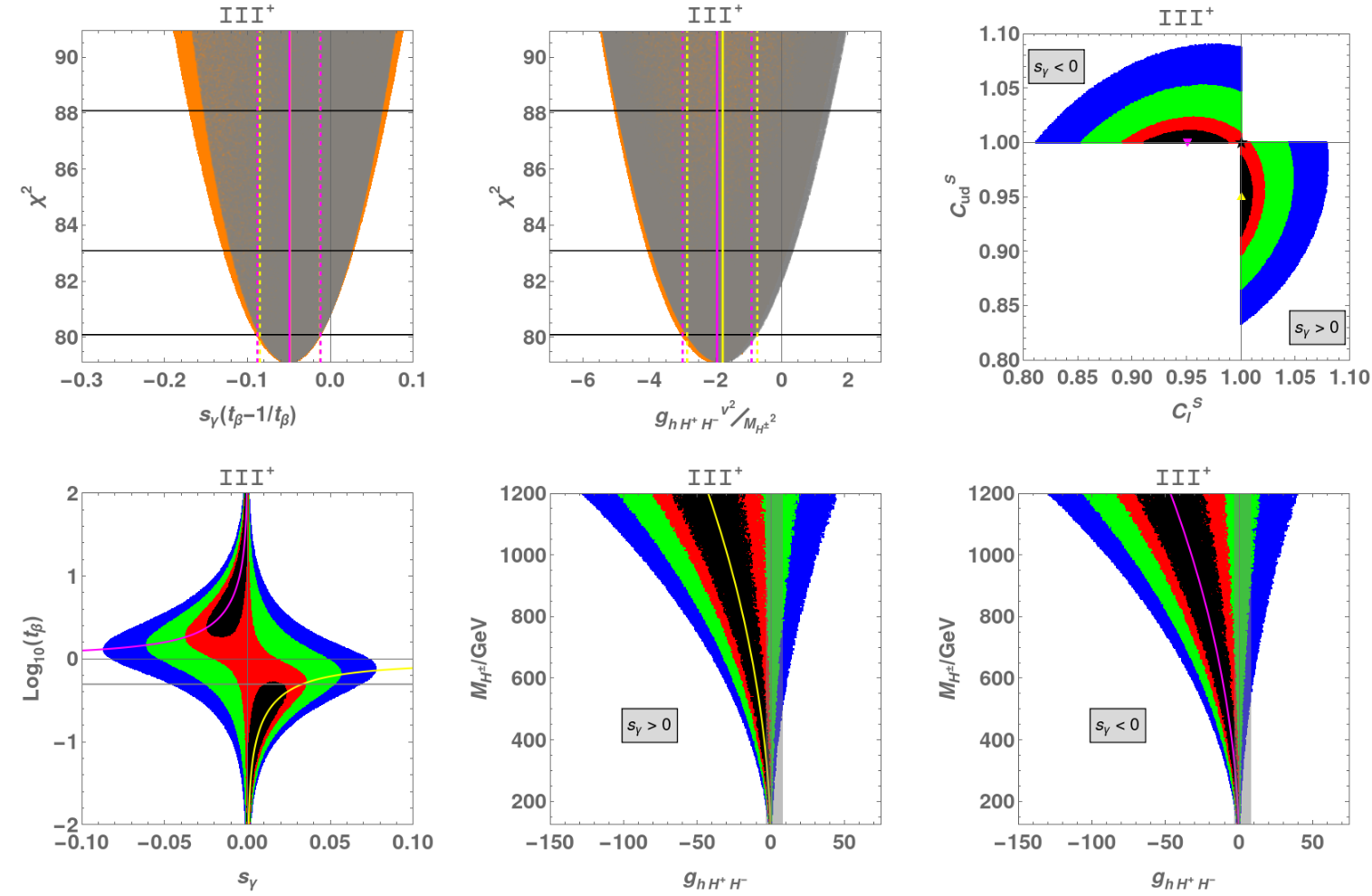}
\end{center}
\vspace{-0.5cm}
\caption{\it {\bf III}$^+$:
The same as in Fig.~\ref{fig:t2p}, but for the {\bf III}$^+$ scenario.
}
\label{fig:t3p}
\end{figure}
\begin{figure}[!htb]
\begin{center}
\includegraphics[width=18.5cm]{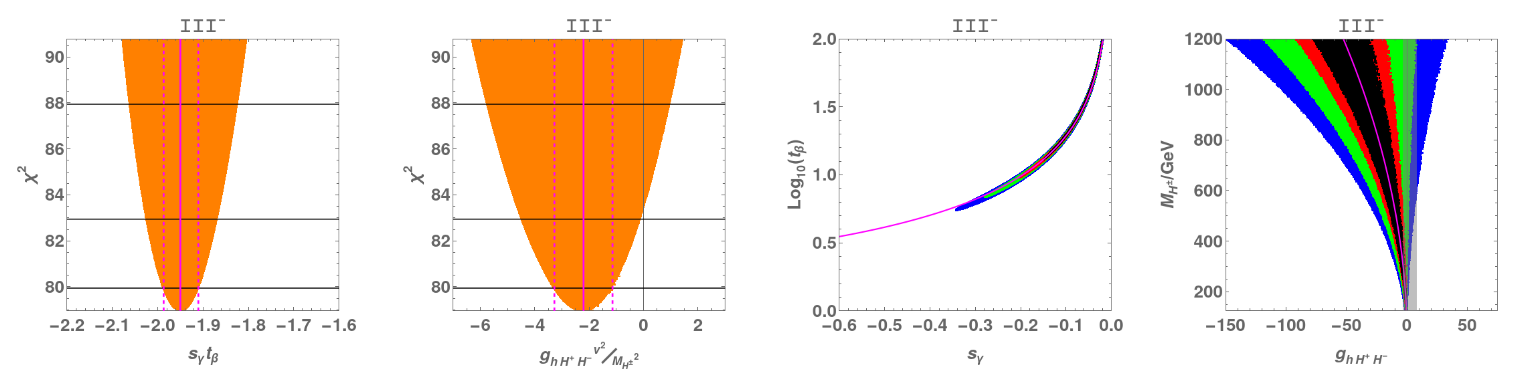}
\end{center}
\vspace{-0.5cm}
\caption{\it {\bf III}$^-$:
The same as in Fig.~\ref{fig:t2m}, but for the {\bf III}$^-$ scenario.
}
\label{fig:t3m}
\end{figure}
Figure~\ref{fig:t3p} shows the results for {\bf III}$^+$, where we observe a two-fold accidental
degeneracy depending on the sign of $s_\gamma$, as in {\bf II}$^+$ and:
\begin{itemize}
\item
$\chi^2_{\rm min}/{\rm dof}=79.11/73\,(79.10/73)$
and ${\rm gof}=0.2923\,(0.2923)$ for $s_\gamma>0\,(s_\gamma<0)$
\item Parameteric relations at the minima:
$s_\gamma(t_\beta-1/t_\beta) \simeq -0.05$ and
$g_{_{hH^+H^-}}(v^2/M_{H^\pm}^2) \simeq -2.0$
\item $1\sigma$ CIs:
\begin{itemize}[label={$\circ$}]
\item $s_\gamma>0$:
$[0\,,0.02]$ for $s_\gamma$,
$[-0.09\,,-0.01]$ for $s_\gamma(t_\beta-1/t_\beta)$, and
$[-2.9\,,-0.8]$ for $g_{_{hH^+H^-}}(v^2/M_{H^\pm}^2)$
\item $s_\gamma<0$:
$[-0.02\,,0]$ for $s_\gamma$,
$[-0.09\,,-0.01]$ for $s_\gamma(t_\beta-1/t_\beta)$, and
$[-3.0\,,-0.9]$ for $g_{_{hH^+H^-}}(v^2/M_{H^\pm}^2)$
\end{itemize}
\item Characteristic features:
The two degenerate minima are clearly
separated in the $(C_{ud}^S\,,C_{\ell}^S)$ plane, with the best-fit values of
$(C_{ud}^S\,,C_{\ell}^S) \simeq (0.95\,,1)$ for $s_\gamma>0$ and
$(C_{ud}^S\,,C_{\ell}^S) \simeq (1\,,0.95)$ for $s_\gamma<0$
\item Imposing $t_\beta>1/2$ and the UBE conditions:
$(\chi^2_{\rm min})_{\rm Constrained}^{{\bf III}^+}/{\rm dof}=79.13/73$ and
$({\rm gof})_{\rm Constrained}^{{\bf III}^+}=0.2916$, which are similar to
the results without constraints
\end{itemize}

Figure~\ref{fig:t3m} is for {\bf III}$^-$ for which we find:
\begin{itemize}
\item
$\chi^2_{\rm min}/{\rm dof}=78.96/73$
and ${\rm gof}=0.2962$
\item Parameteric relations at the minima:
$s_\gamma t_\beta \simeq -1.95$ and
$g_{_{hH^+H^-}}(v^2/M_{H^\pm}^2) \simeq -2.2$
\item $1\sigma$ CIs:
$[-0.15\,,-0.02]$ for $s_\gamma$,
$[-1.99\,,-1.91]$ for $s_\gamma t_\beta$, and
$[-3.3\,,-1.2]$ for $g_{_{hH^+H^-}}(v^2/M_{H^\pm}^2)$
\item Imposing $t_\beta>1/2$ and the UBE conditions:
$(\chi^2_{\rm min})_{\rm Constrained}^{{\bf III}^-}/{\rm dof}=78.97/73$ and
$({\rm gof})_{\rm Constrained}^{{\bf III}^-}=0.2961$, which are similar to
the results without constraints
\end{itemize}

\begin{figure}[!htb]
\begin{center}
\includegraphics[width=14.0cm]{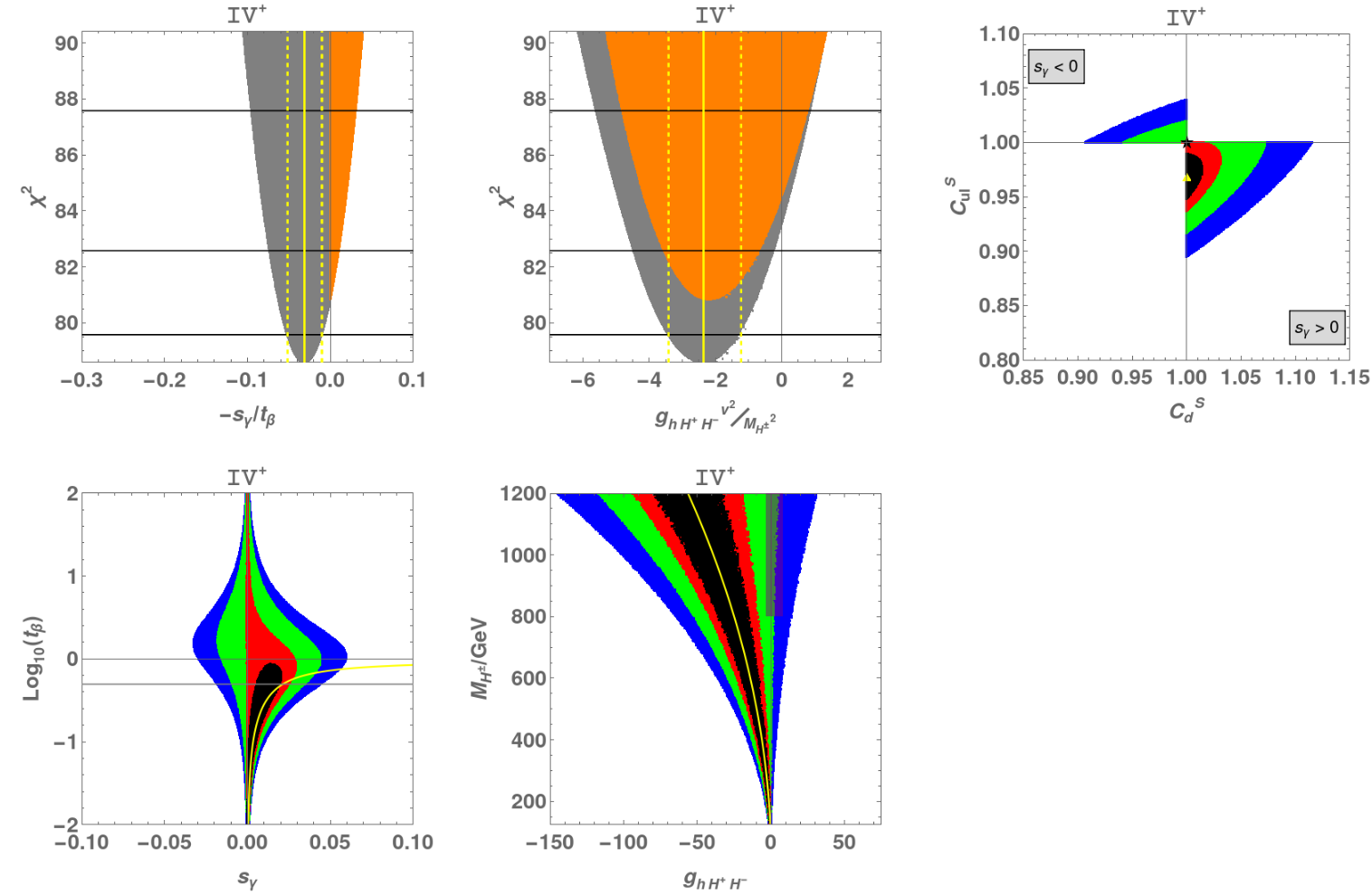}
\end{center}
\vspace{-0.5cm}
\caption{\it {\bf IV}$^+$:
The same as in Fig.~\ref{fig:t2p}, but for the {\bf IV}$^+$ scenario.
}
\label{fig:t4p}
\end{figure}
\begin{figure}[!htb]
\begin{center}
\includegraphics[width=18.5cm]{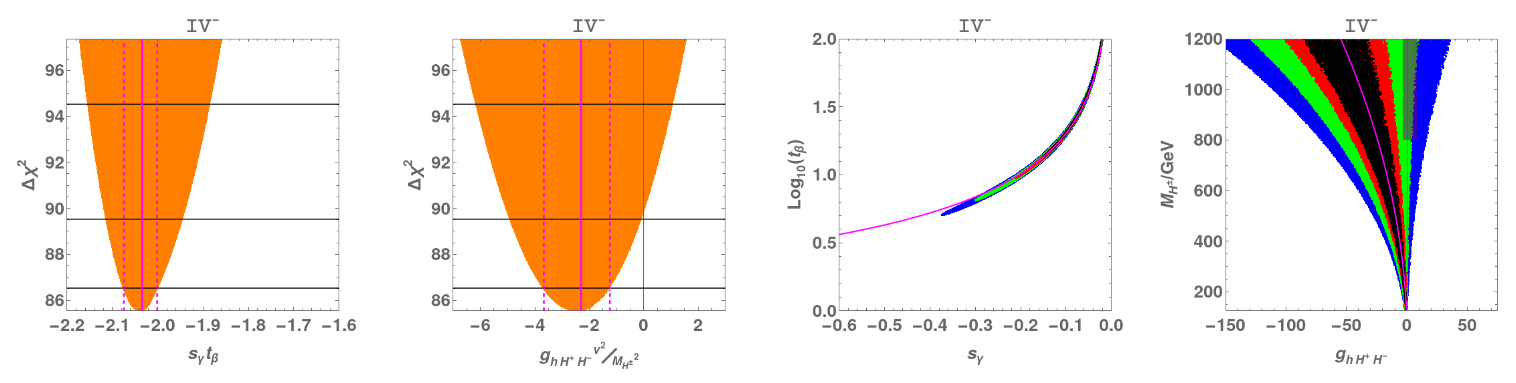}
\end{center}
\vspace{-0.5cm}
\caption{\it {\bf IV}$^-$:
The same as in Fig.~\ref{fig:t2m}, but for the {\bf IV}$^-$ scenario.
}
\label{fig:t4m}
\end{figure}

Figure~\ref{fig:t4p} is for {\bf IV}$^+$ for which we find:
\begin{itemize}
\item
$\chi^2_{\rm min}/{\rm dof}=78.60/73$
and ${\rm gof}=0.3062$
\item Parameteric relations at the minima:
$-s_\gamma/t_\beta \simeq -0.03$ and
$g_{_{hH^+H^-}}(v^2/M_{H^\pm}^2) \simeq -2.4$
\item $1\sigma$ CIs:
$[0\,,0.02]$ for $s_\gamma$,
$[-0.05\,,-0.01]$ for $-s_\gamma/t_\beta$, and
$[-3.4\,,-1.2]$ for $g_{_{hH^+H^-}}(v^2/M_{H^\pm}^2)$
\item Characteristic features: $s_\gamma>0$ at 68\% CL
\item Imposing $t_\beta>1/2$, the UBE conditions, and $M_{H^\pm}>800$ GeV:
$(\chi^2_{\rm min})_{\rm Constrained}^{{\bf IV}^+}/{\rm dof}=83.50/73$ and
$({\rm gof})_{\rm Constrained}^{{\bf IV}^+}=0.1880$,
which are quite worse
compared to the results without constraints
\end{itemize}

Figure~\ref{fig:t4m} is for {\bf IV}$^-$ for which we find:
\begin{itemize}
\item
$\chi^2_{\rm min}/{\rm dof}=85.55/73$
and ${\rm gof}=0.1495$
\item Parameteric relations at the minima:
$s_\gamma t_\beta \simeq -2.04$ and
$g_{_{hH^+H^-}}(v^2/M_{H^\pm}^2) \simeq -2.3$
\item $1\sigma$ CIs:
$[-0.15\,,-0.02]$ for $s_\gamma$,
$[-2.08\,,-2.00]$ for $s_\gamma t_\beta$, and
$[-3.7\,,-1.3]$ for $g_{_{hH^+H^-}}(v^2/M_{H^\pm}^2)$
\item Imposing $t_\beta>1/2$, the UBE conditions, and $M_{H^\pm}>800$ GeV:
$(\chi^2_{\rm min})_{\rm Constrained}^{{\bf IV}^-}/{\rm dof}=89.35/73$ and
$({\rm gof})_{\rm Constrained}^{{\bf IV}^-}=0.0938$,
which are quite worse
compared to the results without constraints
\end{itemize}

\begin{figure}[!htb]
\begin{center}
\includegraphics[width=9.0cm]{app.png}
\includegraphics[width=9.0cm]{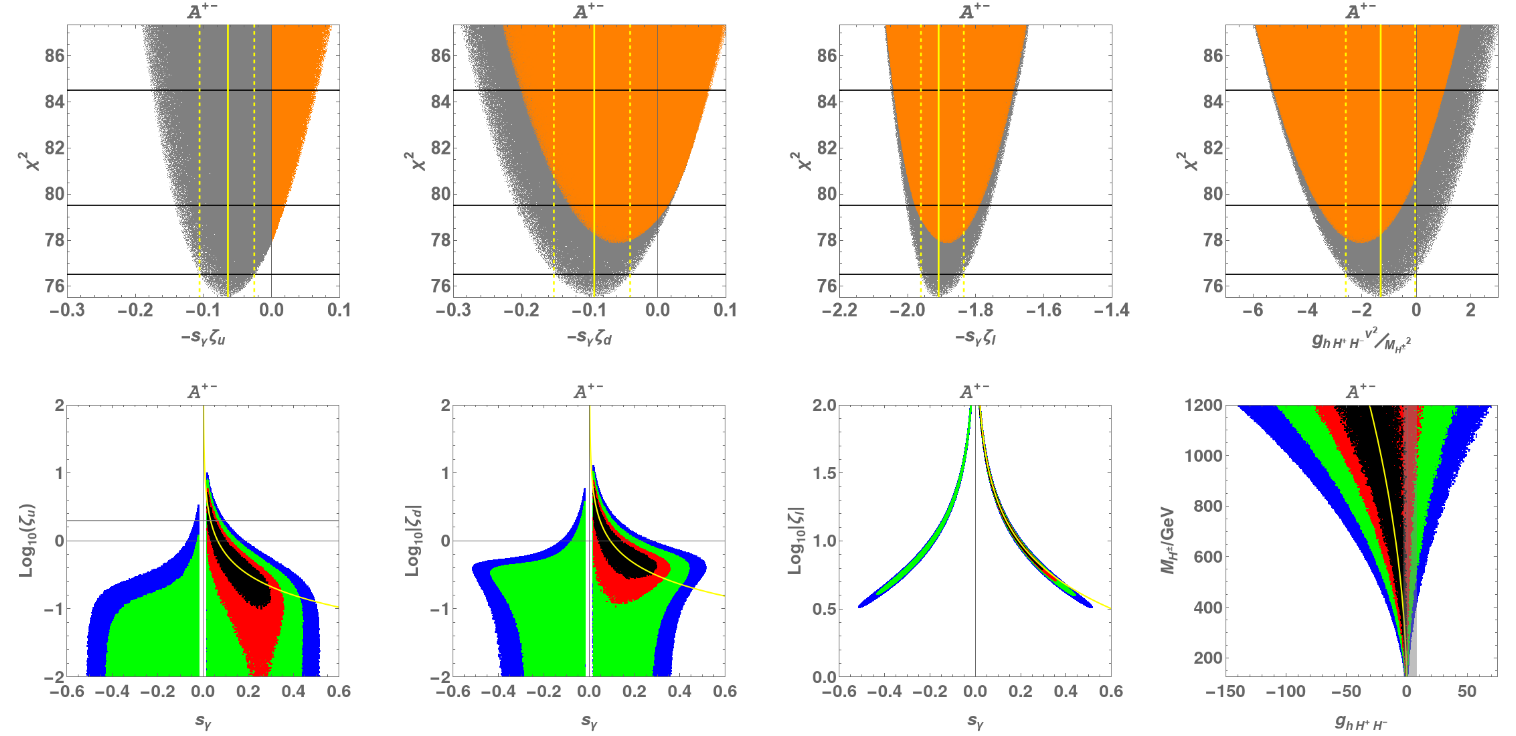}
\includegraphics[width=9.0cm]{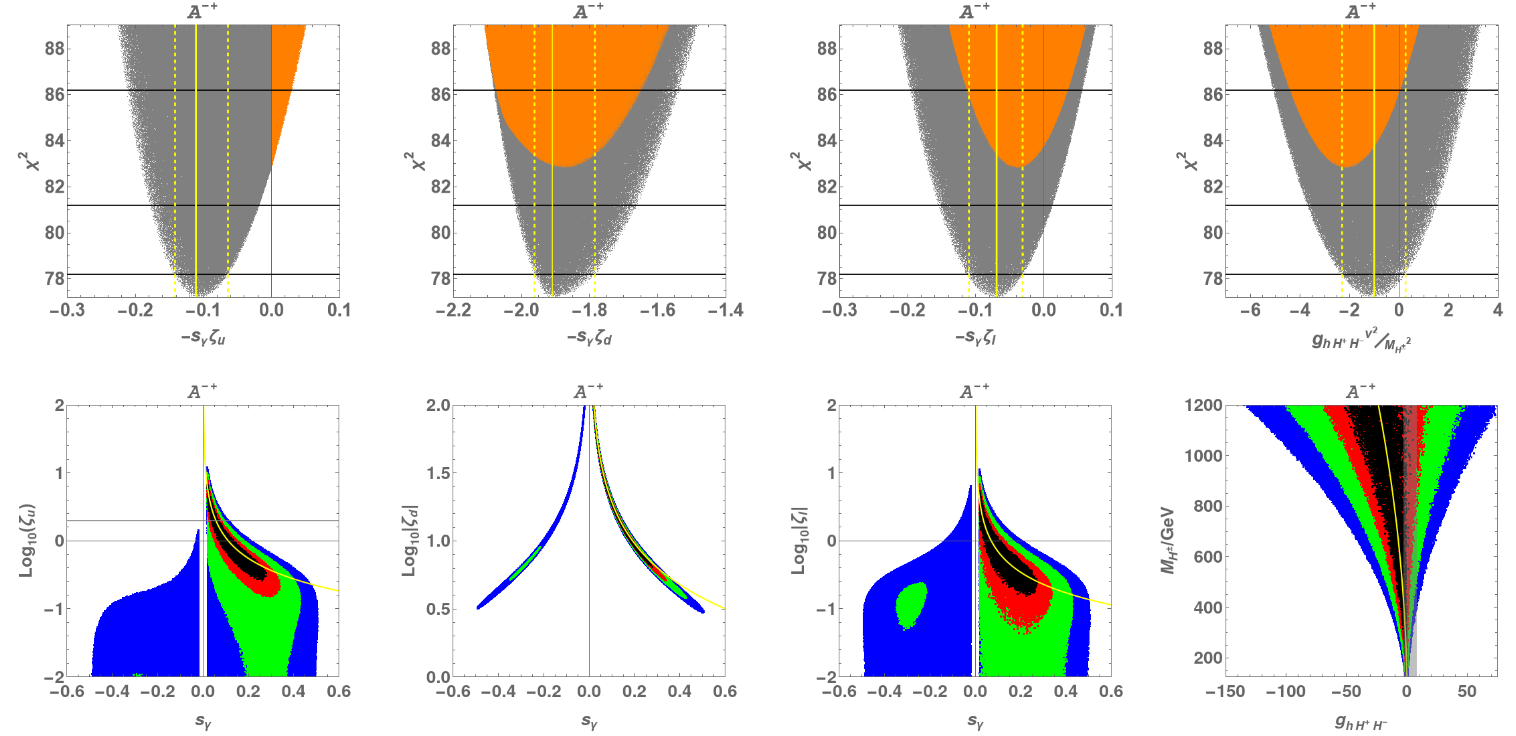}
\includegraphics[width=9.0cm]{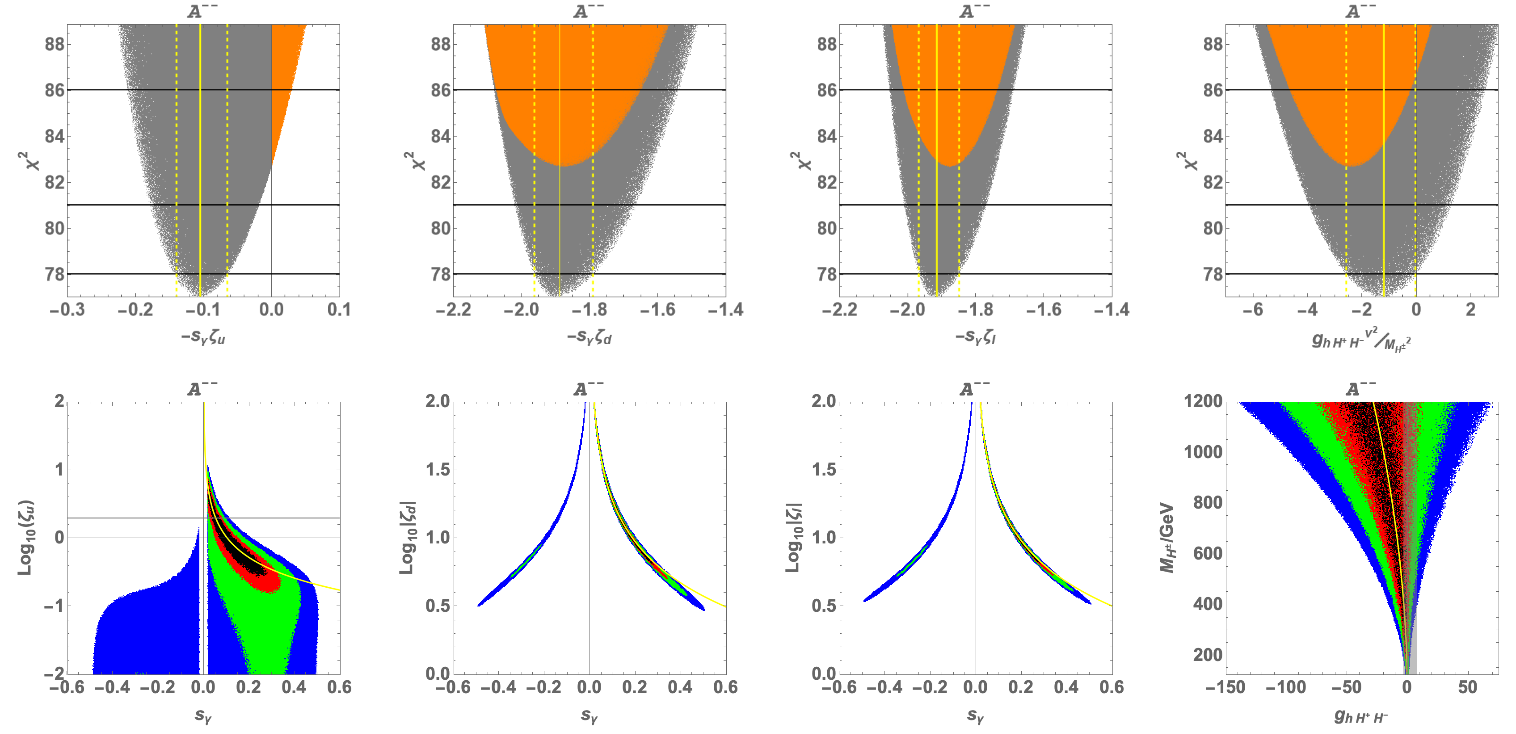}
\end{center}
\vspace{-0.5cm}
\caption{\it
{\bf A}$^{++}$ (upper-left), {\bf A}$^{+-}$ (upper-right),
{\bf A}$^{-+}$ (lower-left), and {\bf A}$^{--}$ (lower-right).
The upper-left panel for {\bf A}$^{++}$
is the same as Fig.~\ref{fig:app}  and
the lines, shades, and colors are the same as in Fig.~\ref{fig:t2p}.
}
\label{fig:Aligned}
\end{figure}

The four panels in Fig.~\ref{fig:Aligned} correspond to
the four Aligned 2HDM scenarios of
{\bf A}$^{++}$ (upper-left), {\bf A}$^{+-}$ (upper-right),
{\bf A}$^{+-}$ (lower-left), and {\bf A}$^{-+}$ (lower-right)
with each panel containing 8 frames. Note that
the upper-left panel for {\bf A}$^{++}$
is the same as Fig.~\ref{fig:app}, and we include it again for comparison
purposes.
Our findings are:
\begin{itemize}
\item $\chi^2_{\rm min}/{\rm dof}=75.70/71\,({\bf A}^{++})$,
$75.52/71\,({\bf A}^{+-})$,
$77.20/71\,({\bf A}^{-+})$, and
$77.03/71\,({\bf A}^{--})$
\item ${\rm gof}=0.3302\,({\bf A}^{++})$,
$0.3347\,({\bf A}^{+-})$,
$0.2873\,({\bf A}^{-+})$, and
$0.2918\,({\bf A}^{--})$
\item Parametric relations at the minima
\begin{itemize}[label={$\circ$}]
\item {\bf A}$^{++}$:
$-s_\gamma\zeta_u\simeq -0.07 $,
$-s_\gamma\zeta_d\simeq -0.09 $,
$-s_\gamma\zeta_\ell\simeq -0.07 $, and
$g_{_{hH^+H^-}}(v^2/M_{H^\pm}^2)\simeq -1.1$
\item {\bf A}$^{+-}$:
$-s_\gamma\zeta_u\simeq -0.06 $,
$-s_\gamma\zeta_d\simeq -0.09 $,
$-s_\gamma\zeta_\ell\simeq -1.91 $, and
$g_{_{hH^+H^-}}(v^2/M_{H^\pm}^2)\simeq -1.3 $
\item {\bf A}$^{-+}$:
$-s_\gamma\zeta_u\simeq -0.11 $,
$-s_\gamma\zeta_d\simeq -1.91 $,
$-s_\gamma\zeta_\ell\simeq -0.07 $, and
$g_{_{hH^+H^-}}(v^2/M_{H^\pm}^2)\simeq -1.0$
\item {\bf A}$^{--}$:
$-s_\gamma\zeta_u\simeq -0.11 $,
$-s_\gamma\zeta_d\simeq -1.89 $,
$-s_\gamma\zeta_\ell\simeq -1.91 $, and
$g_{_{hH^+H^-}}(v^2/M_{H^\pm}^2)\simeq -1.2 $
\end{itemize}
\item $1\sigma$ CIs:
\begin{itemize}[label={$\circ$}]
\item $s_\gamma$:
$[0\,, 0.30 ]\,({\bf A}^{++})$,
$[0.02\,, 0.30 ]\,({\bf A}^{+-})$,
$[0.02\,, 0.28 ]\,({\bf A}^{-+})$, and
$[0.02\,, 0.28 ]\,({\bf A}^{--})$
\item $-s_\gamma\zeta_u$:
$[-0.11\,, -0.03 ]\,({\bf A}^{++})$,
$[-0.11\,, -0.03 ]\,({\bf A}^{+-})$,
$[-0.14\,, -0.06 ]\,({\bf A}^{-+})$, and
$[-0.14\,, -0.07 ]\,({\bf A}^{--})$
\item $-s_\gamma\zeta_d$:
$[-0.15\,, -0.04 ]\,({\bf A}^{++})$,
$[-0.15\,, -0.04 ]\,({\bf A}^{+-})$,
$[-1.96\,, -1.79 ]\,({\bf A}^{-+})$, and
$[-1.96\,, -1.79 ]\,({\bf A}^{--})$
\item $-s_\gamma\zeta_\ell$:
$[-0.12\,, -0.03 ]\,({\bf A}^{++})$,
$[-1.96\,, -1.84 ]\,({\bf A}^{+-})$,
$[-0.11\,, -0.03 ]\,({\bf A}^{-+})$, and
$[-1.97\,, -1.85 ]\,({\bf A}^{--})$
\item $g_{_{hH^+H^-}}(v^2/M_{H^\pm}^2)$:
$[-2.4\,, 0.2 ]\,({\bf A}^{++})$,
$[-2.6\,, -0 ]\,({\bf A}^{+-})$,
$[-2.3\,, 0.3 ]\,({\bf A}^{-+})$, and
$[-2.6\,, -0 ]\,({\bf A}^{--})$
\end{itemize}
\item Characteristic features:
In the wrong-sign scenarios of {\bf A}$^{+-,-+,--}$,
the region of $|s_\gamma|\lsim 0.02$ is not accessible.
This is because
the alignment parameters of $|\zeta_{d,\ell}|$
are scanned up to $100$, while
$-s_\gamma\zeta_{d\,,\ell} \simeq -2$ is required
\item Imposing $\zeta_u<2$ and the UBE conditions:
\footnote{See Table~\ref{tab:results_constrained}.}
\begin{itemize}[label={$\circ$}]
\item $(\chi^2_{\rm min}/{\rm dof})_{\rm Constrained}$:
$75.99/71\,({\bf A}^{++})$,
$75.57/71\,({\bf A}^{+-})$,
$77.32/71\,({\bf A}^{-+})$, and
$77.12/71\,({\bf A}^{--})$
\item $({\rm gof})_{\rm Constrained}$:
$0.3211\,({\bf A}^{++})$,
$0.3331\,({\bf A}^{+-})$,
$0.2841\,({\bf A}^{-+})$, and
$0.2894\,({\bf A}^{--})$
\end{itemize}
\end{itemize}

\setcounter{equation}{0}
\section{Correlations among couplings}
\label{app:B}
In this Appendix, we present the CL regions in the two-coupling planes
to illustrate the correlations among the 125 GeV
Higgs boson couplings to SM particles in the 12 scenarios of
the six types of 2HDMs considered in this work.
\begin{figure}[!htb]
\begin{center}
\includegraphics[width=9.0cm]{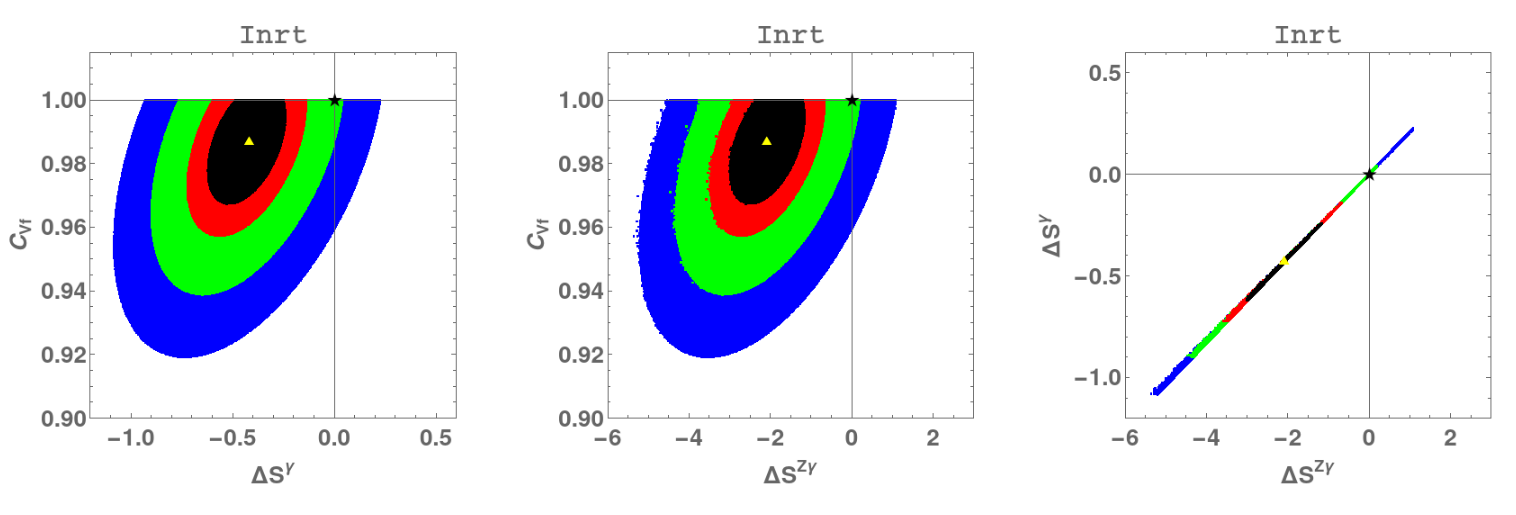}
\includegraphics[width=9.0cm]{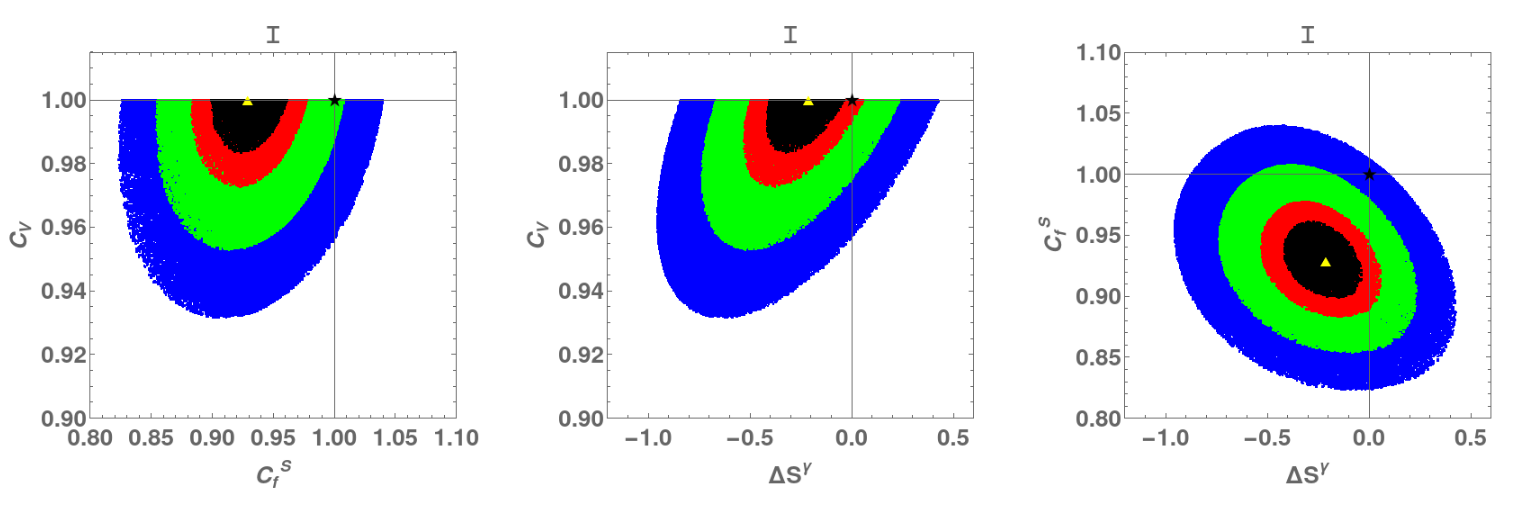}
\end{center}
\vspace{-0.5cm}
\caption{\it
The CL regions of {\bf Inrt} [Left] and {\bf I} [Right]:
In the left panel, we show the CL regions of {\bf Inrt} in the
$(\Delta S^\gamma\,,C_{Vf})$,
$(\Delta S^{Z\gamma}\,,C_{Vf})$, and
$(\Delta S^{Z\gamma}\,,\Delta S^\gamma)$ planes.
In the right panel, we show the CL regions of {\bf I} in the
$(C_f^S\,,C_V)$,
$(\Delta S^\gamma\,,C_V)$, and
$(\Delta S^\gamma\,,C_f^S)$.
The contour regions shown are for
$\Delta\chi^2\leq 1$ (black),
$\Delta\chi^2\leq 2.3$ (red),
$\Delta\chi^2\leq 5.99$ (green),
$\Delta\chi^2\leq 11.83$ (blue)
above the minimum, which correspond to
confidence levels of 39.35\%, 68.27\%, 95\%, and 99.73\%, respectively.
In each frame, the vertical and horizontal lines indicate the SM point, denoted
by a star, while the best-fit point is marked by a triangle.
}
\label{fig:t01_corr}
\end{figure}

\begin{figure}[!htb]
\begin{center}
\includegraphics[width=9.0cm]{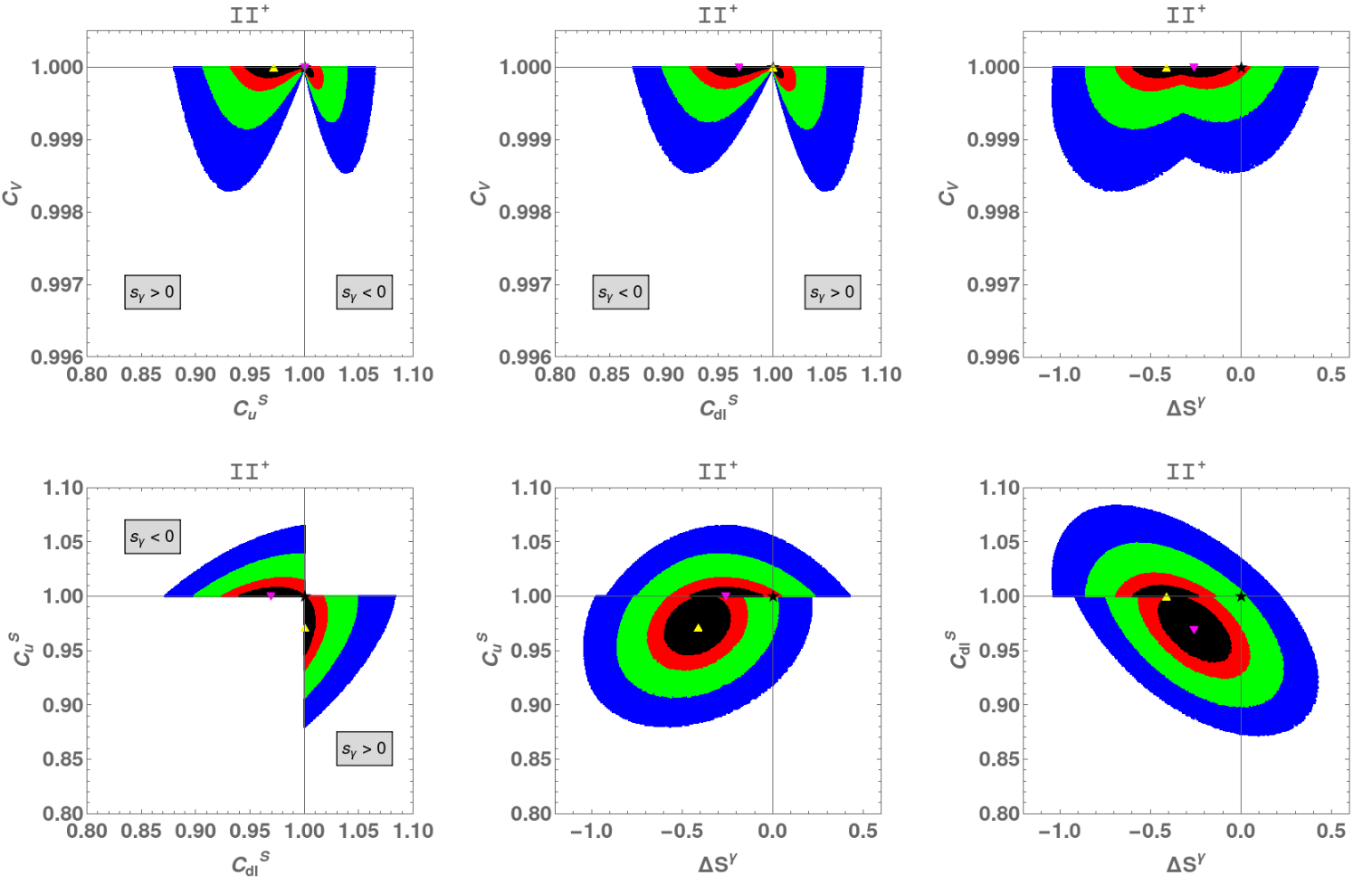}
\includegraphics[width=9.0cm]{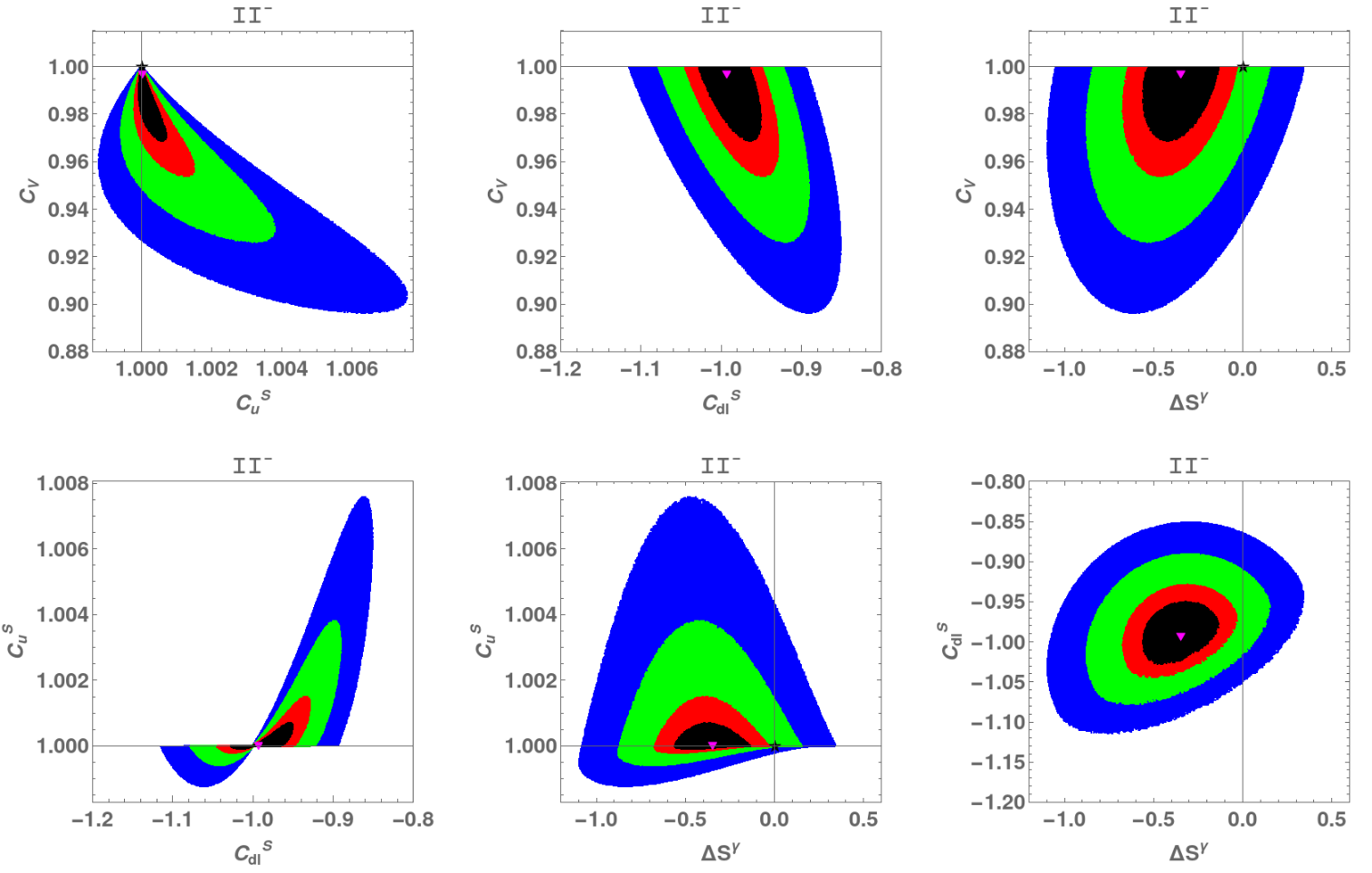}
\end{center}
\vspace{-0.5cm}
\caption{\it
The CL regions of {\bf II}$^+$ [Left] and {\bf II}$^-$ [Right]:
In each panel, the CL regions are shown in the
$(C_u^S\,,C_V)$ (upper-left),
$(C_{d\ell}^S\,,C_V)$ (upper-middle),
$(\Delta S^\gamma\,,C_V)$ (upper-right),
$(C_{d\ell}^S\,,C_u^S)$ (lower-left),
$(\Delta S^\gamma\,,C_u^S)$ (lower-middle), and
$(\Delta S^\gamma\,,C_{d\ell}^S)$ (lower-middle) planes.
The lines, colors, and markers are the same as in Fig.~\ref{fig:t01_corr}.
}
\label{fig:t2_corr}
\end{figure}

\begin{figure}[!htb]
\begin{center}
\includegraphics[width=9.0cm]{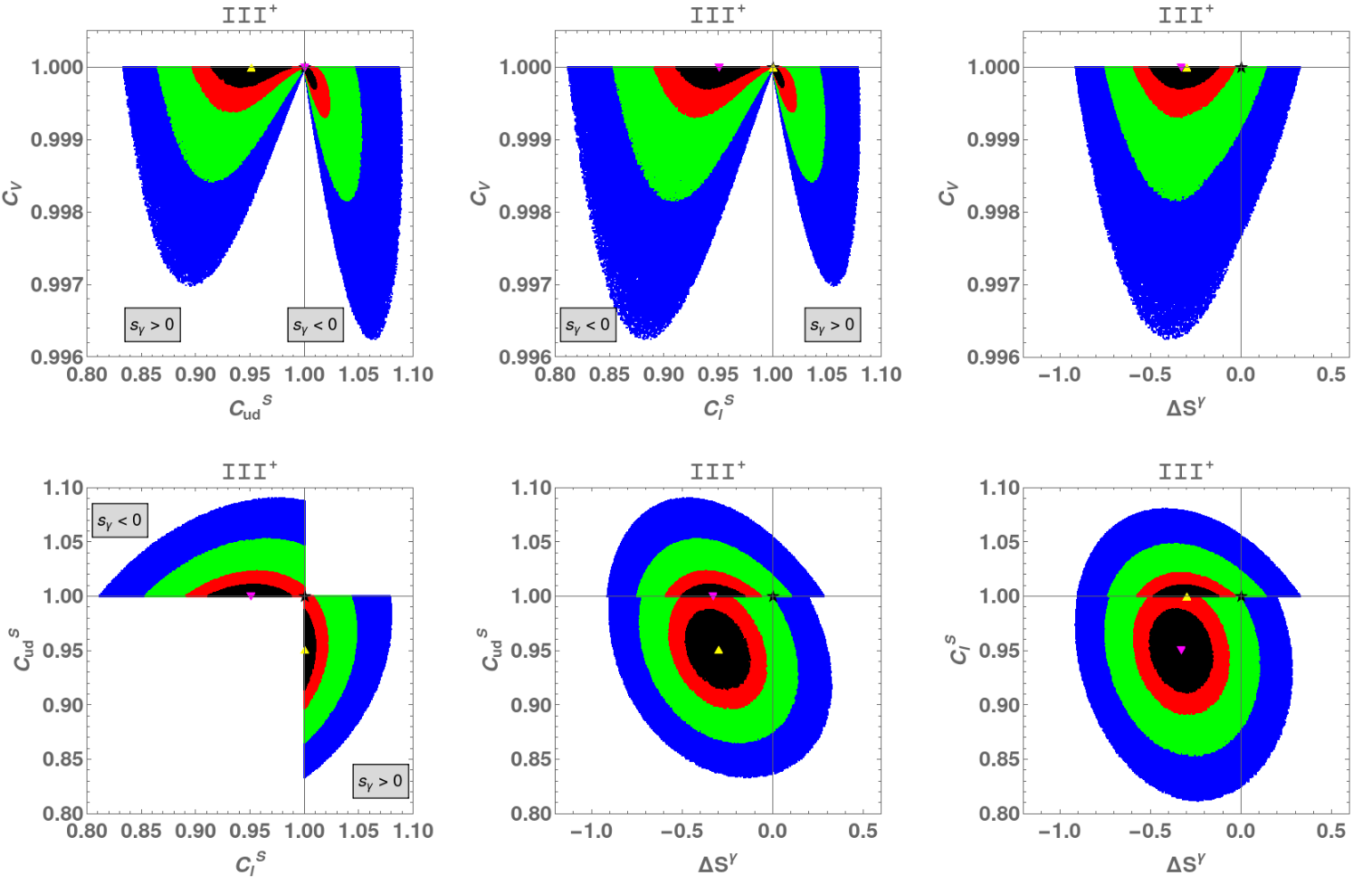}
\includegraphics[width=9.0cm]{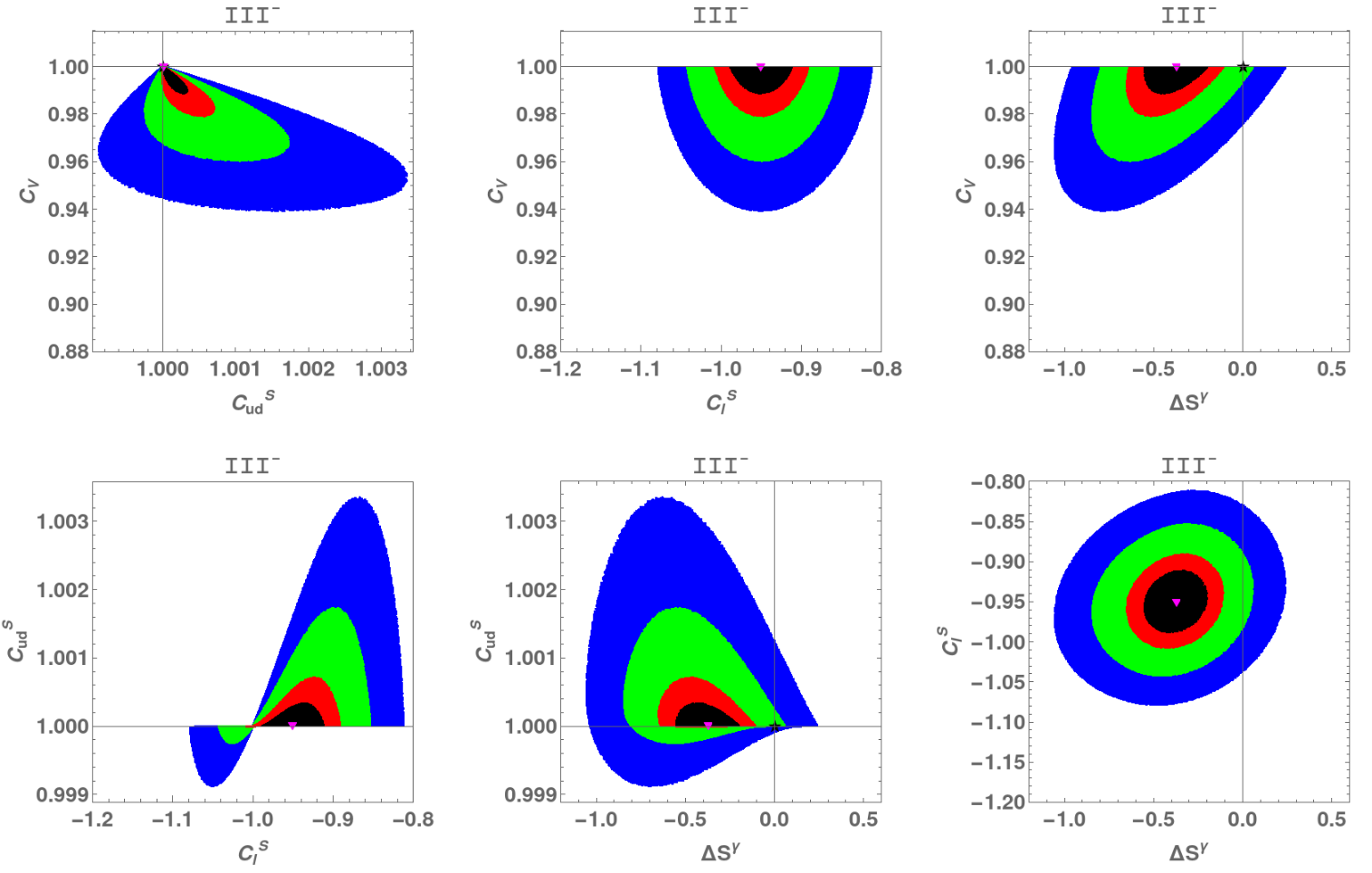}
\end{center}
\vspace{-0.5cm}
\caption{\it
The CL regions of {\bf III}$^+$ [Left] and {\bf III}$^-$ [Right]:
In each panel, the CL regions are shown in the
$(C_{ud}^S\,,C_V)$ (upper-left),
$(C_{\ell}^S\,,C_V)$ (upper-middle),
$(\Delta S^\gamma\,,C_V)$ (upper-right),
$(C_{\ell}^S\,,C_{ud}^S)$ (lower-left),
$(\Delta S^\gamma\,,C_{ud}^S)$ (lower-middle), and
$(\Delta S^\gamma\,,C_{\ell}^S)$ (lower-middle) planes.
The lines, colors, and markers are the same as in Fig.~\ref{fig:t01_corr}.
}
\label{fig:t3_corr}
\end{figure}

\begin{figure}[!htb]
\begin{center}
\includegraphics[width=9.0cm]{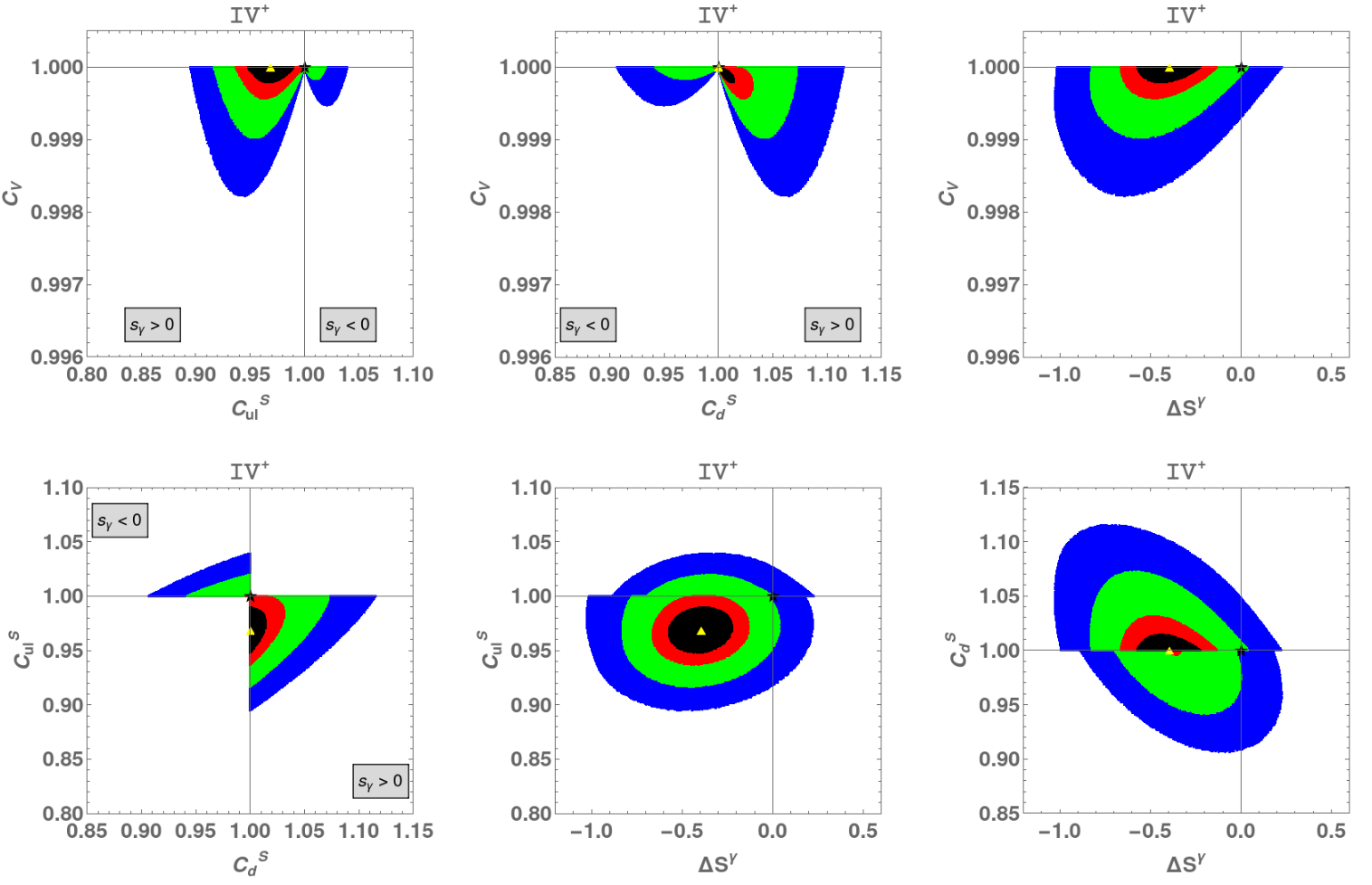}
\includegraphics[width=9.0cm]{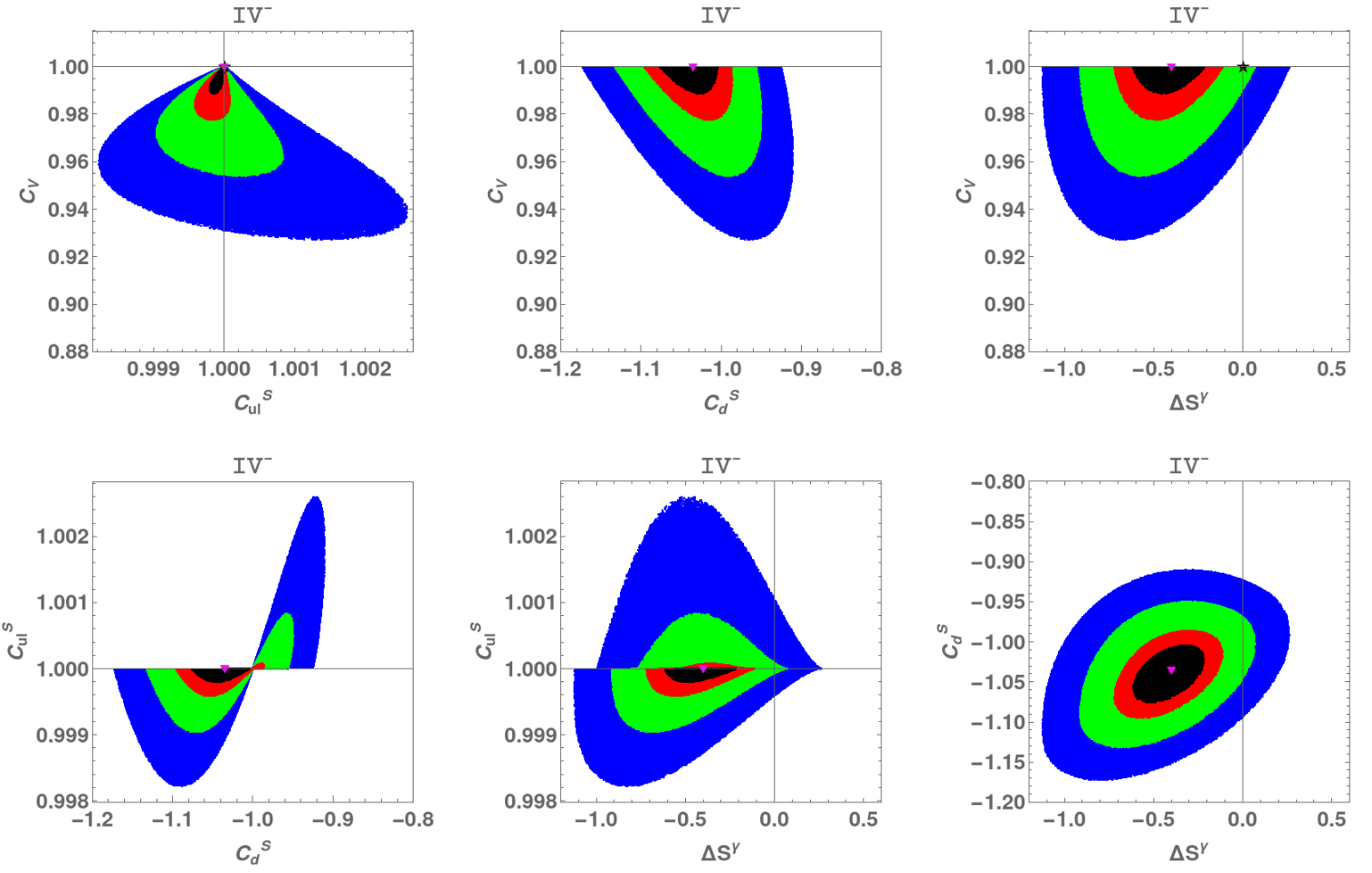}
\end{center}
\vspace{-0.5cm}
\caption{\it
The CL regions of {\bf IV}$^+$ [Left] and {\bf IV}$^-$ [Right]:
In each panel, the CL regions are shown in the
$(C_{u\ell}^S\,,C_V)$ (upper-left),
$(C_{d}^S\,,C_V)$ (upper-middle),
$(\Delta S^\gamma\,,C_V)$ (upper-right),
$(C_{d}^S\,,C_{u\ell}^S)$ (lower-left),
$(\Delta S^\gamma\,,C_{u\ell}^S)$ (lower-middle), and
$(\Delta S^\gamma\,,C_{d}^S)$ (lower-middle) planes.
The lines, colors, and markers are the same as in Fig.~\ref{fig:t01_corr}.
}
\label{fig:t4_corr}
\end{figure}

\begin{figure}[!htb]
\begin{center}
\includegraphics[width=9.0cm]{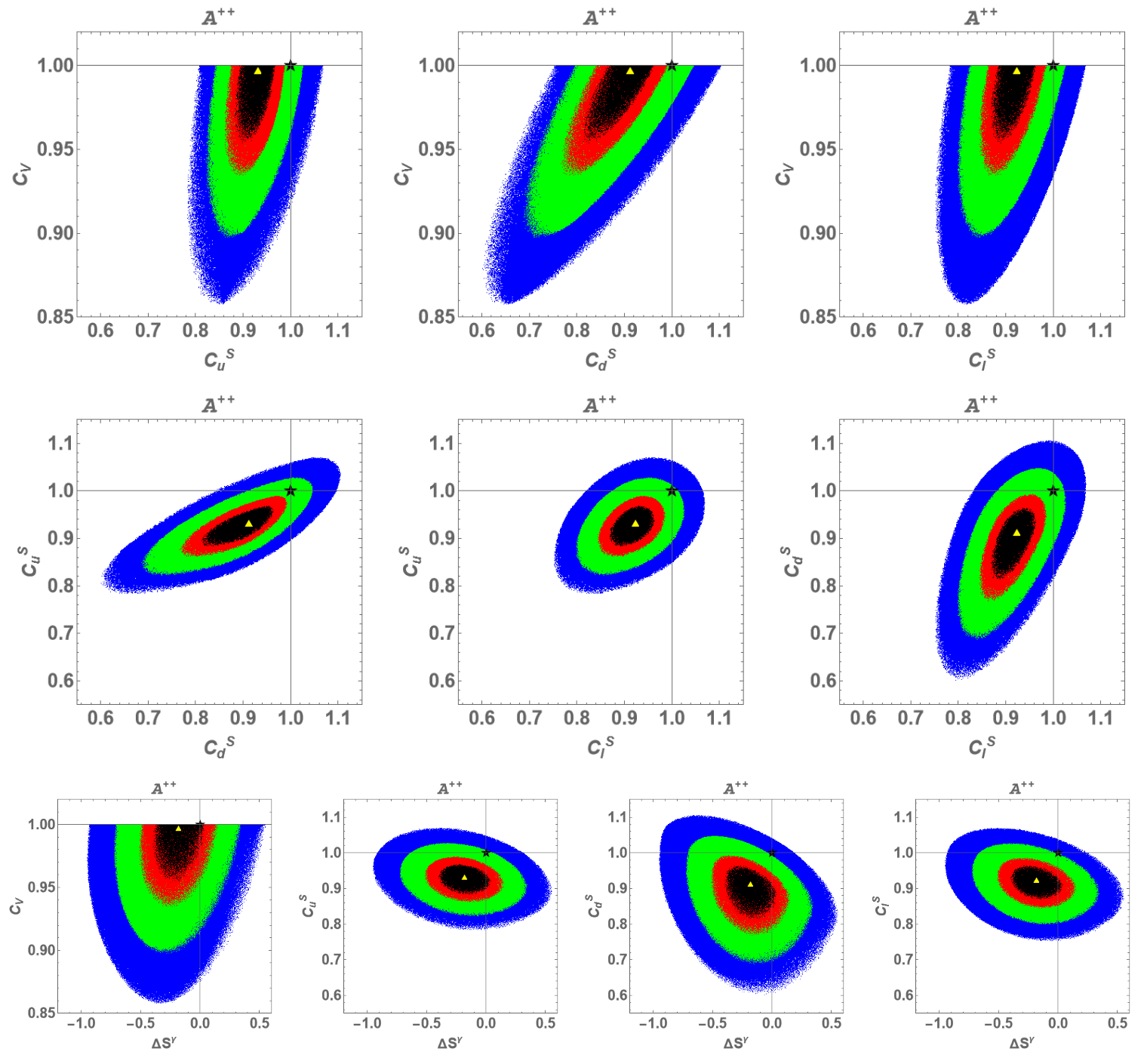}
\includegraphics[width=9.0cm]{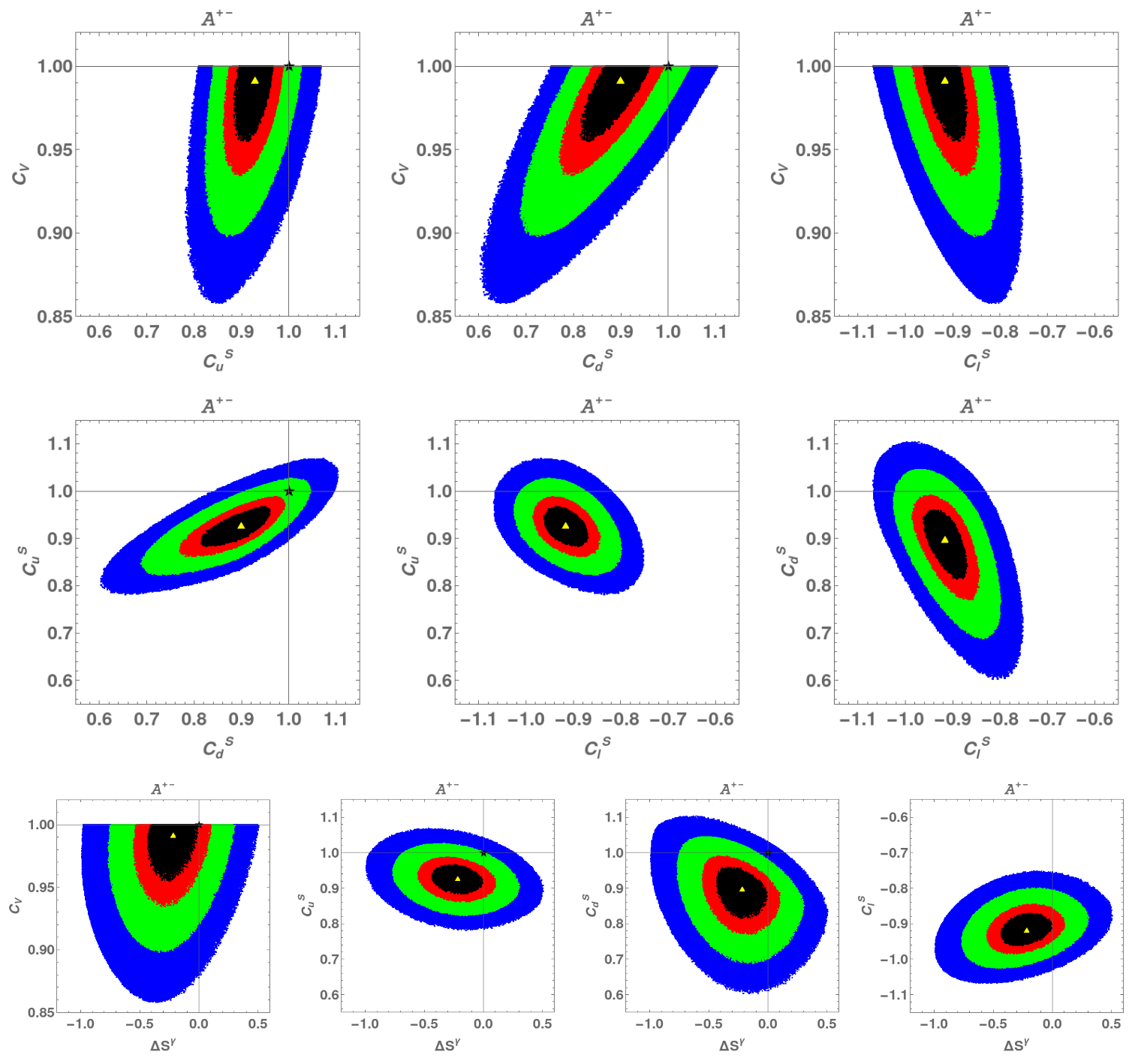}
\includegraphics[width=9.0cm]{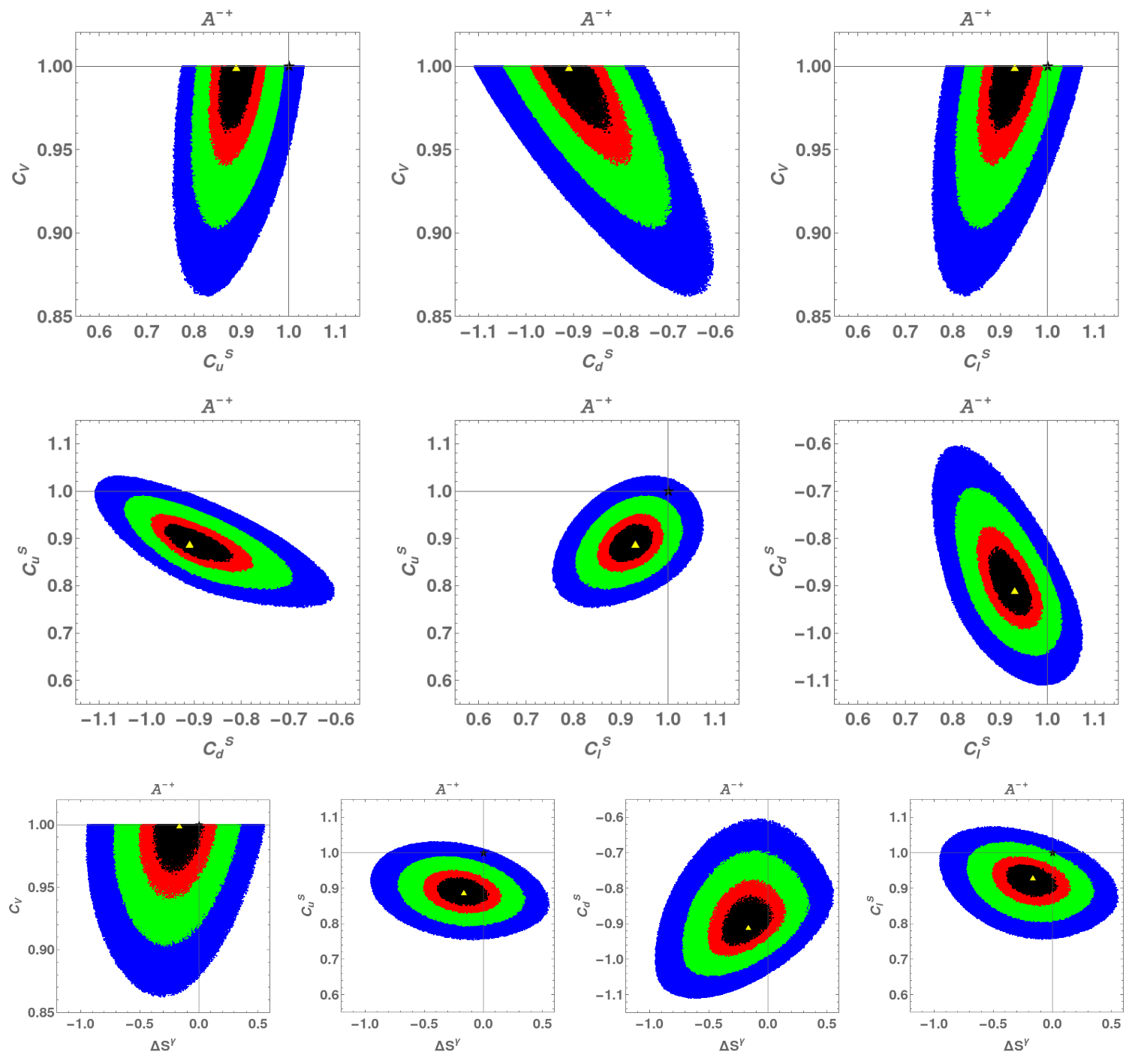}
\includegraphics[width=9.0cm]{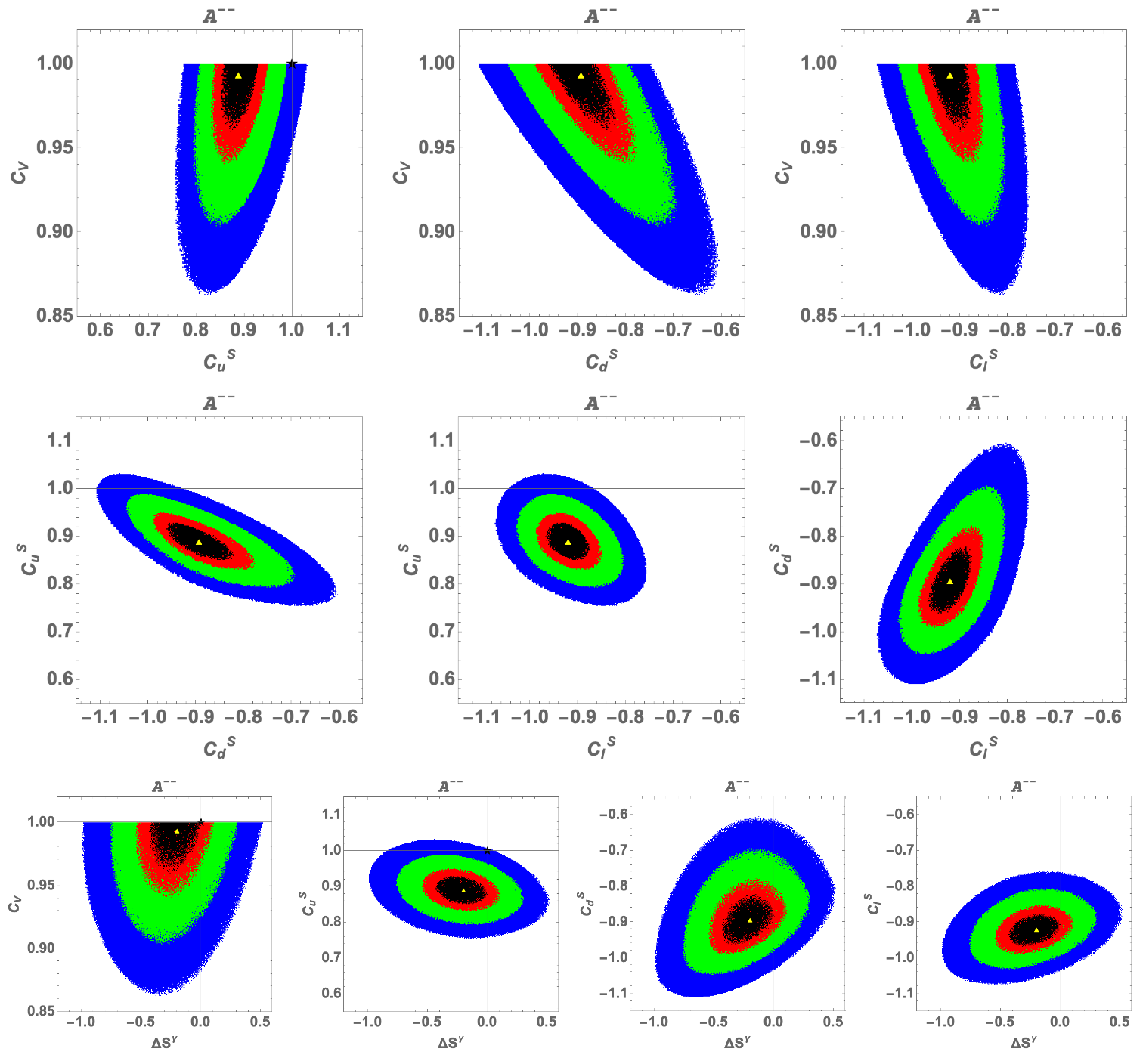}
\end{center}
\vspace{-0.5cm}
\caption{\it
The CL regions of
{\bf A}$^{++}$ [Upper-Left],
{\bf A}$^{+-}$ [Upper-Right],
{\bf A}$^{-+}$ [Lower-Left], and
{\bf A}$^{--}$ [Lower-Right]:
In each panel, the CL regions are shown in the
$(C_{u}^S\,,C_V)$ (upper-left),
$(C_{d}^S\,,C_V)$ (upper-middle),
$(C_{\ell}^S\,,C_V)$ (upper-right),
$(C_{d}^S\,,C_u^S)$ (middle-left),
$(C_{\ell}^S\,,C_u^S)$ (middle-middle),
$(C_{\ell}^S\,,C_d^S)$ (middle-right),
$(\Delta S^\gamma\,,C_V)$ (lower-left),
$(\Delta S^\gamma\,,C_{u}^S)$ (lower-middle-left),
$(\Delta S^\gamma\,,C_{d}^S)$ (lower-middle-right),
$(\Delta S^\gamma\,,C_{\ell}^S)$ (lower-right) planes.
The lines, colors, and markers are the same as in Fig.~\ref{fig:t01_corr}.
}
\label{fig:Aligned_corr}
\end{figure}

\medskip
We highlight the following key points regarding the correlations in the
two-dimensional planes of the Higgs couplings:
\begin{itemize}
\item {\bf Inrt}:
In the left panel of Fig.~\ref{fig:t01_corr}, the CL regions of {\bf Inrt} in the
$(\Delta S^\gamma\,,C_{Vf})$,
$(\Delta S^{Z\gamma}\,,C_{Vf})$, and
$(\Delta S^{Z\gamma}\,,\Delta S^\gamma)$ planes are shown. We observe that
the correlation between $C_{Vf}$ and $\Delta S^{Z\gamma}$ (middle) can be
easily inferred from the correlation between $C_{Vf}$ and $\Delta S^{\gamma}$ (left),
following the relation $\Delta S^{Z\gamma}\simeq 5\Delta S^{\gamma}$,
as clearly shown in the right frame
in the $(\Delta S^{Z\gamma}\,,\Delta S^\gamma)$ plane.
In the remaining scenarios from now on, therefore, we drop the correlations
involved with $\Delta S^{Z\gamma}$, as they can be easily deduced from those
with $\Delta S^{\gamma}$.
\item {\bf I} (Right panel of Fig.~\ref{fig:t01_corr}):
In this best-fit scenario, the SM point lies outside
the 95\% CL region in the $(\Delta S^{\gamma}\,,C_f^S)$ plane, as shown in
the right frame.
\item {\bf II}$^{+}$ (Left panel of Fig.~\ref{fig:t2_corr}):
The lower-left frame is the same as the upper-right
one in Fig.~\ref{fig:t2p}. The two separate CL regions,
depending ${\rm sign}(s_\gamma)$
(also appearing in the upper-left and upper-middle frames),
might essentially explain the discontinuities seen in
the lower-middle and lower-right frames.
The SM point locates at the boundary between the
68\% and 95\% CL regions in the $(\Delta S^\gamma\,,C_{d\ell}^S)$ plane,
as shown in the lower-right frame.
Note that the two degenerate minima are marked by the
yellow ($s_\gamma>0)$ and magenta ($s_\gamma<0)$ triangles.
\item {\bf II}$^{-}$ (Right panel of Fig.~\ref{fig:t2_corr}):
The deviation of $C_u^S$ from 1 is below 1\% at 99.7\% CL.
\item {\bf III}$^{+}$ (Left panel of Fig.~\ref{fig:t3_corr}):
The lower-left frame is the same as the upper-right
one in Fig.~\ref{fig:t3p}, and the similar observations to those made in {\bf II}$^+$
can be applied here.
\item {\bf III}$^{-}$ (Right panel of Fig.~\ref{fig:t3_corr}):
The deviation of $C_{ud}^S$ from 1 is below $0.4$\% at 99.7\% CL.
\item {\bf IV}$^{+}$ (Left panel of Fig.~\ref{fig:t4_corr}):
The SM points locate at the boundary between the
68\% and 95\% CL regions in the
$(\Delta S^\gamma\,,C_V)$ and $(\Delta S^\gamma\,,C_{d}^S)$ planes,
as shown in  the upper-right and lower-right frames.
\item {\bf IV}$^{-}$ (Right panel of Fig.~\ref{fig:t4_corr}):
The deviation of $C_{u\ell}^S$ from 1 is below $0.3$\% at 99.7\% CL.
\item {\bf A}$^{++}$ (Upper-Left panel of Fig.~\ref{fig:Aligned_corr}):
The 68\% CL regions (and most of 95\% CL regions)
lie where all of the Yukawa couplings are smaller than 1,
as shown in the three middle frames. This is consistent with our
finding that the type-I 2HDM gives the best fit.
The SM points locate at the boundary between the
68\% and 95\% CL regions in the
$(\Delta S^\gamma\,,C_u^S)$, $(\Delta S^\gamma\,,C_{d}^S)$,
and $(\Delta S^\gamma\,,C_{\ell}^S)$ planes, as shown in the lower frames.
\item {\bf A}$^{++,+-,-+,--}$ (Fig.~\ref{fig:Aligned_corr}):
We find a stronger correlation between $C_V$ and $C_d^S$
than between $C_V$ and $C_{u,\ell}^S$, as seen in the upper frames in
each panel.
Again in these upper frames, we observe that
the correlation between $C_V$ and $C_d^S$ flips its sign under
{\bf A}$^{+\pm}$ $\leftrightarrow$ {\bf A}$^{-\pm}$,
while that between $C_V$ and $C_\ell^S$ under
{\bf A}$^{\pm+}$ $\leftrightarrow$ {\bf A}$^{\pm-}$.
Similar sign flips of the correlations are observed across the panels.
\end{itemize}

\setcounter{equation}{0}
\section{Constrained fitting results}
\label{app:C}
%
\begin{table}[!thb]
\caption{\label{tab:results_constrained}
The minimal chi-square per degree of freedom ($\chi^2_{\rm min}$/dof),
goodness of fit (gof),
$1\sigma$ confidence interval of $s_\gamma$,
the 95\% CL region of $g_{_{hH^+H^-}}$,
and the best-fitted values
of the  125 GeV Higgs couplings to SM particles
in the 12 scenarios considered in our global fit to the six types of 2HDMs.
Imposed are the constraints from:
the UBE conditions for {\bf Inrt};
$t_\beta>1/2$ and the UBE conditions for {\bf I} and {\bf III}$^\pm$;
$t_\beta>1/2$, the UBE conditions, and $M_{H^\pm}>800$ GeV for
{\bf II}$^\pm$ and {\bf IV}$^\pm$; and
$\zeta_u<2$ and the UBE conditions for {\bf A}$^{\pm\pm,\pm\mp}$.
For the SM, we obtain
$\chi^2_{\rm min}/{\rm dof}=85.29/77$ and ${\rm gof}=0.2424$.
}  \vspace{1mm}
\renewcommand{\arraystretch}{1.3}
\begin{adjustbox}{width=\textwidth}
$\begin{array}{|c|c||c|c||c|c||c|cccccc|c|c|}
\hline
\multicolumn{2}{|c||}{} &  \chi^2_\textrm{min}/\textrm{dof} & \textrm{gof} & s_\gamma & g_{_{hH^+H^-}} & C_{V} & \multicolumn{6}{c|}{C_{f}^{S} } & \Delta S^{\gamma} & \Delta S^{Z\gamma}  \\ \hline
\multirow{2}{*}{\textbf{Inrt}} & s_\gamma>0 &
\multirow{2}{*}{80.28/74} & \multirow{2}{*}{0.2889} & [0,\, 0.25] &
\multirow{2}{*}{$[-3.2,\, 8.4]$} &
\multicolumn{7}{c|}{\multirow{2}{*}{ $C_{Vf}= 0.9872^{+0.0128}_{-0.0198}$ }} &
\multirow{2}{*}{$-0.421^{+0.180}_{-0.194}$} & \multirow{2}{*}{$-2.100^{+0.926}_{-0.910}$} \\ \cline{2-2} \cline{5-5}
 & s_\gamma<0  & & & [-0.25, 0] & & \multicolumn{7}{c|}{} & &
\\ \hline
\multicolumn{2}{|c||}{\textbf{I} } & 76.03/73 & 0.3810 & [0.03, 0.19] & [-3.0,\, 8.3] & 0.9967^{+0.0030}_{-0.0143} & \multicolumn{6}{c|}{ 0.928^{+0.031}_{-0.027} } & -0.231^{+0.193}_{-0.191} & -1.119^{+0.938}_{-0.956} \\ \hline
\multicolumn{2}{|c||}{\textbf{II}^+ } & 81.12/73 & 0.2410 & [-0.02, 0] & [-2.5,\, 8.1] & 1.0_{-0.0001} & \multicolumn{3}{c|}{ C_{u}^{S} = 1.0^{+0.006}_{-0.000} } & \multicolumn{3}{c|}{ C_{d\ell}^S = 0.950^{+0.027}_{-0.022} } & -0.031^{+0.079}_{-0.006} & -0.149^{+0.380}_{-0.030} \\ \hline
\multicolumn{2}{|c||}{\textbf{II}^- } & 89.07/73 & 0.0972 & [-0.19, -0.02] & [-3.1,\, 8.4] & 0.9988^{+0.0010}_{-0.0170} & \multicolumn{3}{c|}{C_{u}^{S} = 1.0^{+0.001}_{-0.000} } & \multicolumn{3}{c|}{C_{d\ell}^S = -0.975^{+0.033}_{-0.020} } & -0.026^{+0.062}_{-0.014} & -0.126^{+0.297}_{-0.066} \\ \hline
\multicolumn{2}{|c||}{\textbf{III}^+ } & 79.13/73 & 0.2916 & [-0.02, 0] & [-2.5,\, 8.1] & 1.0_{-0.0003} & \multicolumn{3}{c|}{C_{ud}^{S} = 1.0^{+0.010}_{-0.000} } & \multicolumn{3}{c|}{C_{\ell}^{S} = 0.944^{+0.039}_{-0.029} } & -0.303^{+0.150}_{-0.185} & -1.487^{+0.747}_{-0.909} \\ \hline
\multicolumn{2}{|c||}{\textbf{III}^- } & 78.97/73 & 0.2961 & [-0.14, -0.02] & [-3.0,\, 8.3] & 0.9997^{+0.0001}_{-0.0100} & \multicolumn{3}{c|}{C_{ud}^{S} = 1.0^{+0.0003}_{-0.000} } & \multicolumn{3}{c|}{C_{\ell}^{S} = -0.947^{+0.030}_{-0.040} } & -0.383^{+0.175}_{-0.147} & -1.859^{+0.841}_{-0.722} \\ \hline
\multicolumn{2}{|c||}{\textbf{IV}^+ } & 83.50/73 & 0.1880 & [-0.01, 0.02] & [-2.5,\, 8.0] & 1.0^{}_{-0.0001} & \multicolumn{3}{c|}{C_{u\ell}^{S} = 0.982^{+0.020}_{-0.018} } & \multicolumn{3}{c|}{C_{d}^{S} = 1.004^{+0.007}_{-0.048} } & -0.031^{+0.038}_{-0.006} & -0.148^{+0.181}_{-0.029} \\ \hline
\multicolumn{2}{|c||}{\textbf{IV}^- } & 89.35/73 & 0.0938 & [-0.11, -0.02] & [-2.9,\, 8.3] & 0.9996^{+0.0002}_{-0.0060} & \multicolumn{3}{c|}{C_{u\ell}^{S} = 1.0^{+0.000}_{-0.0001} } & \multicolumn{3}{c|}{C_{d}^{S} = -1.007^{+0.027}_{-0.031} } & -0.029^{+0.047}_{-0.009} & -0.138^{+0.226}_{-0.042} \\ \hline
\multicolumn{7}{c|}{} &\multicolumn{2}{c|}{ C_{u}^{S} } &\multicolumn{2}{c|}{ C_{d}^{S} } & \multicolumn{2}{c|}{ C_{\ell}^{S} } & \multicolumn{2}{c}{ }  \\ \hline
\multicolumn{2}{|c||}{\textbf{A}^{++} } & 75.99/71 & 0.3211 & [0.03, 0.27] & [-3.4,\, 8.4] & 0.9983^{+0.0011}_{-0.0364} & \multicolumn{2}{c|}{ 0.922^{+0.044}_{-0.032} } &\multicolumn{2}{c|}{ 0.887^{+0.069}_{-0.074} } & \multicolumn{2}{c|}{ 0.911^{+0.045}_{-0.040} } & -0.074^{+0.141}_{-0.311} & -0.355^{+0.680}_{-1.520} \\ \hline
\multicolumn{2}{|c||}{\textbf{A}^{+-} } & 75.57/71 & 0.3331 & [0.03, 0.27] & [-3.4,\, 8.4] & 0.9913^{+0.0083}_{-0.0283} & \multicolumn{2}{c|}{ 0.932^{+0.020}_{-0.045} } &\multicolumn{2}{c|}{ 0.910^{+0.037}_{-0.080} } & \multicolumn{2}{c|}{ -0.924^{+0.045}_{-0.030} } & -0.240^{+0.233}_{-0.149} & -1.164^{+1.131}_{-0.727} \\ \hline
\multicolumn{2}{|c||}{\textbf{A}^{-+} } & 77.32/71 & 0.2841 & [0.05, 0.27] & [-3.4,\, 8.4] & 0.9879^{+0.0111}_{-0.0253} & \multicolumn{2}{c|}{ 0.890^{+0.032}_{-0.035} } &\multicolumn{2}{c|}{ -0.887^{+0.063}_{-0.059} } & \multicolumn{2}{c|}{ 0.909^{+0.045}_{-0.034} } & -0.142^{+0.188}_{-0.245} & -0.685^{+0.907}_{-1.198} \\ \hline
\multicolumn{2}{|c||}{\textbf{A}^{--} } & 77.12/71 & 0.2894 & [0.05, 0.26] & [-3.4,\, 8.4] & 0.9968^{+0.0018}_{-0.0309} & \multicolumn{2}{c|}{ 0.884^{+0.046}_{-0.029} } &\multicolumn{2}{c|}{ -0.899^{+0.068}_{-0.043} } & \multicolumn{2}{c|}{ -0.931^{+0.047}_{-0.026} } & -0.182^{+0.181}_{-0.245} & -0.879^{+0.876}_{-1.194} \\ \hline
\end{array}$
\end{adjustbox}
\end{table}
\begin{figure}[!htb]
\begin{center}
\includegraphics[width=16.5cm]{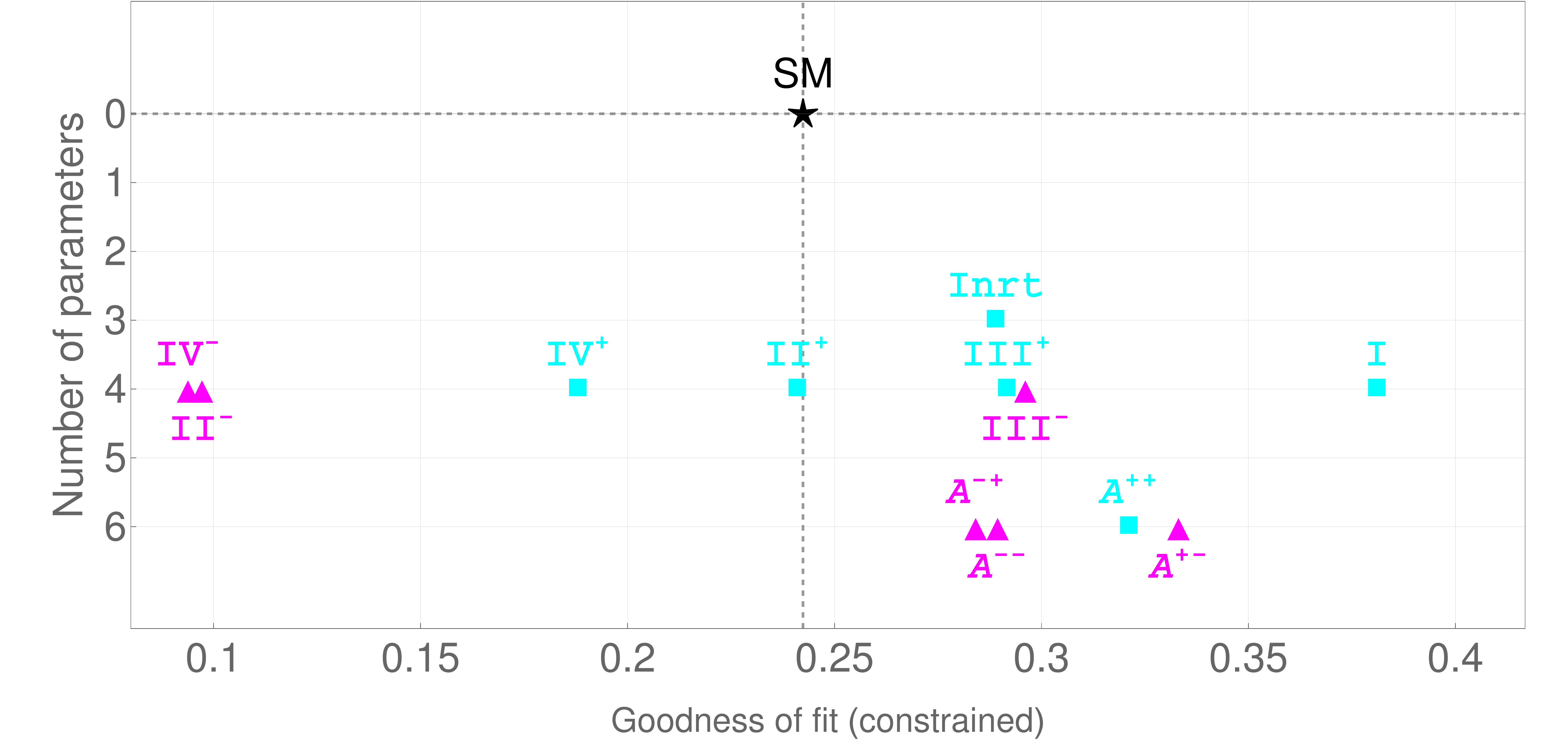}
\end{center}
\vspace{-0.5cm}
\caption{\it
The same as in Fig.~\ref{fig:gof}, but imposing
the constraints from:
the UBE conditions for {\bf Inrt};
$t_\beta>1/2$ and the UBE conditions for {\bf I} and {\bf III}$^\pm$;
$t_\beta>1/2$, the UBE conditions, and $M_{H^\pm}>800$ GeV for
{\bf II}$^\pm$ and {\bf IV}$^\pm$; and
$\zeta_u<2$ and the UBE conditions for {\bf A}$^{\pm\pm,\pm\mp}$.
}
\label{fig:gof_constrained}
\end{figure}
In this Appendix, we show the fitting results obtained by imposing
the following constraints depending on scenarios:
\begin{itemize}
\item {\bf Inrt}: the UBE conditions
\item {\bf I} and {\bf III}$^\pm$: $t_\beta>1/2$ and the UBE conditions
\item {\bf II}$^\pm$ and {\bf IV}$^\pm$:
$t_\beta>1/2$, the UBE conditions, and $M_{H^\pm}>800$ GeV~\cite{Misiak:2020vlo}
\item {\bf A}$^{\pm\pm,\pm\mp}$:
$\zeta_u<2$ and the UBE conditions
\end{itemize}
In this work, we have not considered other constraints,
such as those from flavor observables other than the radiative $b\to s\gamma$
decay or direct searches for heavy scalars at the LHC.
This is because we concentrate on the 125 GeV Higgs couplings to
SM particles, without specifying the detailed properties of
the heavier Higgs staes.

\medskip

In Table~\ref{tab:results_constrained} and Fig.~\ref{fig:gof_constrained},
we present the fitting results obtained by imposing the constraints outlined
above.
While two degenerate minima remain indistinguishable in {\bf Inrt}, the
accidental degeneracies in {\bf II}$^+$ and {\bf III}$^+$ are lifted.
Overall, we observe that the constraints from $t_\beta>1/2$ (or $\zeta_u<2$)
and the UBE conditions have little effect on the fitting results by themselves.
However,
when the UBE conditions are combined with $M_{H^\pm}>800$ GeV,
as in the cases of {\bf II}$^\pm$ and {\bf IV}$^\pm$, the fits become significantly worse.
This is obviously because $\Delta S^{Z\gamma}/5\simeq \Delta S^\gamma \simeq
g_{_{hH^+H^-}}(v^2/M_{H^\pm}^2)/6$ cannot deviate from the SM value of 0
when the two constraints are simultaneously imposed and one fails to accommodate
the LHC data $\mu(\sum{\cal P},\gamma\gamma)=1.10\pm 0.07$~\cite{Heo:2024cif} and
$\widehat\mu^{\rm EXP}(Z\gamma)=2.2\pm 0.7$~\cite{ATLAS:2023yqk} comfortably.
Now we have the worse gof values than the SM in {\bf II}$^\pm$ and {\bf IV}$^\pm$.
Otherwise, we again find that
scenario {\bf I} provides the best fit.
The wrong-sign $C_d^S$ leads to the worse fits
in {\bf A}$^{-\pm}$ compared to {\bf A}$^{+\pm}$. Additionally,
the wrong-sign scenarios of {\bf III}$^-$, {\bf A}$^{+-}$, and {\bf A}$^{--}$,
with $C_\ell^S<0$, result in the better fits
than the corresponding ones with $C_\ell^S>0$, though the difference is slight.

\medskip

For the couplings, we observe similar behavior to the unconstrained cases,
except for
$\Delta S^{\gamma}$ and $\Delta S^{Z\gamma}$ in {\bf II}$^\pm$ and {\bf IV}$^\pm$,
which are significantly reduced by imposing
the UBE constraints and $M_{H^\pm}>800$ GeV.
Note that
we also provide the 95\% CL region of $g_{_{hH^+H^-}}$ in
Table~\ref{tab:results_constrained}.

\end{appendix}



\begin{thebibliography}{99}
%

\bibitem{ATLAS:2012yve}
G.~Aad \textit{et al.} [ATLAS],
``Observation of a new particle in the search for the Standard Model Higgs boson
with the
ATLAS detector at the LHC,''
Phys. Lett. B \textbf{716}, 1-29 (2012)
doi:10.1016/j.physletb.2012.08.020
[arXiv:1207.7214 [hep-ex]].

\bibitem{CMS:2012qbp}
S.~Chatrchyan \textit{et al.} [CMS],
``Observation of a New Boson at a Mass of 125 GeV with the CMS Experiment at the
LHC,''
Phys. Lett. B \textbf{716}, 30-61 (2012)
doi:10.1016/j.physletb.2012.08.021
[arXiv:1207.7235 [hep-ex]].

\bibitem{ATLAS:2022vkf}
 [ATLAS],
``A detailed map of Higgs boson interactions by the ATLAS experiment ten years after
the discovery,''
Nature \textbf{607} (2022) no.7917, 52-59
[erratum: Nature \textbf{612} (2022) no.7941, E24]
doi:10.1038/s41586-022-04893-w
[arXiv:2207.00092 [hep-ex]].

\bibitem{CMS:2022dwd}
A.~Tumasyan \textit{et al.} [CMS],
``A portrait of the Higgs boson by the CMS experiment ten years after the discovery,''
Nature \textbf{607} (2022) no.7917, 60-68
doi:10.1038/s41586-022-04892-x
[arXiv:2207.00043 [hep-ex]].


\bibitem{Heo:2024cif}
Y.~Heo, D.~W.~Jung and J.~S.~Lee,
``Higgs boson precision analysis of the full LHC run 1 and run 2 data,''
Phys. Rev. D \textbf{110}, no.1, 013003 (2024)
doi:10.1103/PhysRevD.110.013003
[arXiv:2402.02822 [hep-ph]].

\bibitem{Biekotter:2022ckj}
T.~Biek\"otter and M.~Pierre,
``Higgs-boson visible and invisible constraints on hidden sectors,''
Eur. Phys. J. C \textbf{82}, no.11, 1026 (2022)
doi:10.1140/epjc/s10052-022-10990-x
[arXiv:2208.05505 [hep-ph]].

\bibitem{ATLAS:2016neq}
G.~Aad \textit{et al.} [ATLAS and CMS],
``Measurements of the Higgs boson production and decay rates and constraints on its
couplings from a combined ATLAS and CMS analysis of the LHC pp collision data at $
\sqrt{s}=7 $ and 8 TeV,''
JHEP \textbf{08}, 045 (2016)
doi:10.1007/JHEP08(2016)045
[arXiv:1606.02266 [hep-ex]].

\bibitem{Branco:2011iw}
G.~C.~Branco, P.~M.~Ferreira, L.~Lavoura, M.~N.~Rebelo, M.~Sher and J.~P.~Silva,
``Theory and phenomenology of two-Higgs-doublet models,''
Phys. Rept. \textbf{516}, 1-102 (2012)
doi:10.1016/j.physrep.2012.02.002
[arXiv:1106.0034 [hep-ph]].


\bibitem{Pich:2009sp}
A.~Pich and P.~Tuzon,
``Yukawa Alignment in the Two-Higgs-Doublet Model,''
Phys. Rev. D \textbf{80} (2009), 091702
doi:10.1103/PhysRevD.80.091702
[arXiv:0908.1554 [hep-ph]].

\bibitem{Glashow:1976nt}
  S.~L.~Glashow and S.~Weinberg,
  ``Natural Conservation Laws for Neutral Currents,''
  Phys.\ Rev.\ D {\bf 15} (1977) 1958.

\bibitem{Lee:2021oaj}
J.~S.~Lee and J.~Park,
``Yukawa alignment revisited in the Higgs basis,''
Phys. Rev. D \textbf{106}, no.1, 015023 (2022)
doi:10.1103/PhysRevD.106.015023
[arXiv:2110.03908 [hep-ph]].






\bibitem{Donoghue:1978cj}
J.~F.~Donoghue and L.~F.~Li,
``Properties of Charged Higgs Bosons,''
Phys. Rev. D \textbf{19} (1979), 945
doi:10.1103/PhysRevD.19.945

\bibitem{Georgi:1978ri}
H.~Georgi and D.~V.~Nanopoulos,
``Suppression of Flavor Changing Effects From Neutral Spinless Meson Exchange in Gauge Theories,''
Phys. Lett. B \textbf{82} (1979), 95-96
doi:10.1016/0370-2693(79)90433-7


\bibitem{Choi:2021nql}
S.~Y.~Choi, J.~S.~Lee and J.~Park,
``Decays of Higgs bosons in the Standard Model and beyond,''
Prog. Part. Nucl. Phys. \textbf{120}, 103880 (2021)
doi:10.1016/j.ppnp.2021.103880
[arXiv:2101.12435 [hep-ph]].



\bibitem{tevatron_aa_ww}
A. Juste, in Proceedings of HCP2012, 15 November 2012, Kyoto, Japan,
http://kds.kek.jp/conferenceDisplay.py?confId=9237.

\bibitem{Herner:2016woc}
K.~Herner [CDF and D0],
``Higgs Boson Studies at the Tevatron,''
Nucl. Part. Phys. Proc. \textbf{273-275}, 852-856 (2016)
doi:10.1016/j.nuclphysbps.2015.09.131

\bibitem{ATLAS:2023yqk}
G.~Aad \textit{et al.} [ATLAS and CMS],
``Evidence for the Higgs Boson Decay to a Z Boson and a Photon at the LHC,''
Phys. Rev. Lett. \textbf{132}, 021803 (2024)
doi:10.1103/PhysRevLett.132.021803
[arXiv:2309.03501 [hep-ex]].

\bibitem{Misiak:2020vlo}
M.~Misiak, A.~Rehman and M.~Steinhauser,
``Towards $ \overline{B}\to {X}_s\gamma $ at the NNLO in QCD without interpolation in m$_{c}$,''
JHEP \textbf{06}, 175 (2020)
doi:10.1007/JHEP06(2020)175
[arXiv:2002.01548 [hep-ph]].





\end{thebibliography}
\end{document}